\documentclass[12pt]{article}

\usepackage[utf8]{inputenc}
\usepackage[T1]{fontenc}

\usepackage[body={18cm, 23cm, centered}]{geometry}
\usepackage{amsmath,amssymb,amsthm,amscd,cite,comment,amsfonts,indentfirst,color,setspace,bbold,arydshln}
\usepackage{mathtext,marvosym,textcomp}
\usepackage[makeroom]{cancel}
\usepackage{mathtools}
\usepackage{cancel}

\usepackage[debug,pageanchor=false]{hyperref}
\hypersetup{colorlinks=true,linktocpage,breaklinks,
	urlcolor=blue,
	linkcolor=blue,
	citecolor=blue
}

\usepackage[vcentermath]{youngtab}

\usepackage[mathmode,centertableaux]{ytableau}
\usepackage{tikz}
\usetikzlibrary{arrows}
\usetikzlibrary{shapes.misc}
\usepackage{tikz-cd}

\usetikzlibrary{arrows,shapes,positioning, fit}
\tikzstyle{every picture}+=[remember picture]
\tikzstyle{na} = [baseline=-.5ex]
\tikzstyle{format} = [rounded rectangle,
                      thick,
                      minimum size=1cm,
                      draw=red!00!black!80,
                      top color=white,
                      bottom color=red!00!black!00,
                      on grid]
\tikzstyle{format1} = [rectangle,
                      thick,
                      minimum size=1cm,
                      draw=red!00!black!80,
                      top color=white,
                      bottom color=red!00!black!00,
                      on grid]
\tikzstyle{format0} = [rounded rectangle,
					  thick,
				      minimum size=1.2cm,
					  draw=blue!00!black!80,
					  top color=white,
                      bottom color=blue!00!black!00,
                      text centered]
\tikzstyle{formatd} = [rounded rectangle,
					  thick,
					  dashed,
				      minimum size=1cm,
					  draw=blue!50!black!80,
					  top color=white,
                      bottom color=blue!00!black!00,
                      text centered]
\tikzstyle{format1d} = [rounded rectangle,
                      thick,
                      dashed,
                      minimum size=2cm,
					  draw=blue!50!black!80,
					  top color=white,
                      bottom color=blue!00!black!00,                  
                      text centered]

\tikzset{cross/.style={cross out, draw=black, minimum size=2*(#1-\pgflinewidth), inner sep=0pt, outer sep=0pt},
	cross/.default={5pt}}
\usepackage[ normalem]{ulem}
\usepackage{tocloft}

\numberwithin{equation}{section}

\selectfont

\def\a{\alpha} 
\def\b{\beta} 
\def\g{\gamma} 
\def\d{\delta} 
\def\e{\epsilon}
\def\ve{\varepsilon} 
 
\def\h{\eta} 

\def\k{\kappa} 
\def\l{\lambda} 
\def\m{\mu}
\def\n{\nu} 
\def\x{\xi} 
\def\r{\rho}
\def\q{\theta}
\def\s{\sigma} 
  
\def\f{\phi}
 
\def\w{\omega}

\def\L{\Lambda}

\def\W{\Omega}

\def\fr{\frac}  \def\dt{\partial}

\def\mc{\mathcal}

\def\mF{\mathcal{F}}

\def\mH{\mathcal{H}}
\def\mL{\mathcal{L}}

\def\mH{\mathcal{H}}

\def\mR{\mathcal{R}}
\def\mM{\mathcal{M}}
\def\mZ{\mathcal{Z}}

\def\tx{\tilde{x}}

\def\tdt{\tilde{\partial}}

\def\DD{{\mathcal{D}}}

\def\XX{\mathbb{X}}
\def\RR{\mathbb{R}}
\def\SS{\mathbb{S}}

\def\TT{\mathbb{T}}

\def\sl{\mathfrak{sl}}

\newcommand{\ol}[1]{\overline{#1}}

\def\SL{\mathrm{SL}}

\def\USp{\mathrm{USp}}

\begin{document}
\renewcommand{\refname}{\begin{center}References\end{center}}
	
\begin{titlepage}
		
	\vfill
	\begin{flushright}

	\end{flushright}
		
	\vfill
	
	\begin{center}
		\baselineskip=16pt
		{\Large \bf 
		 Generalized 11D supergravity equations \\ from tri-vector deformations
		}
		\vskip 1cm
		    Ilya Bakhmatov$^{a}$\footnote{\tt ibakhmatov@itmp.msu.ru},
		    Aybike \c{C}atal-\"Ozer$^{b}$\footnote{\tt ozerayb@itu.edu.tr},
		    Nihat Sadik Deger$^{c}$\footnote{\tt sadik.deger@boun.edu.tr},\\
		    Kirill Gubarev$^{d}$\footnote{\tt kirill.gubarev@phystech.edu},
		    Edvard T. Musaev$^{d}$\footnote{\tt musaev.et@phystech.edu},
		\vskip .3cm
		\begin{small}
			{\it
			    $^a$Institute of Theoretical and Mathematical Physics, Moscow State University, Russia, \\
			    $^b$Department of Mathematics, Istanbul Technical University, Istanbul, Turkey, \\
			    $^c$Department of Mathematics, Bogazici University, Bebek, 34342, Istanbul, Turkey, \\
			    $^d$Moscow Institute of Physics and Technology,	Institutskii per. 9, Dolgoprudny, 141700, Russia \\
			}
		\end{small}
	\end{center}
	
	\vfill 
	\begin{center} 
		\textbf{Abstract}
	\end{center} 
	\begin{quote}
	In \href{https://arxiv.org/abs/2203.03372}{arXiv:2203.03372} [Phys.Rev.D 105 (2022) 8, L081904] we presented  a modification of 11-dimensional supergravity field equations which upon dimensional reduction yields generalized supergravity equations in 10-dimensions. In this paper we provide full technical details of that result which is based on SL(5) exceptional field theory. The equations are obtained by making a non-unimodular tri-vector Yang-Baxter deformation which breaks the initial GL(11) symmetry down to GL(7)$\times$GL(4). We also give some non-trivial solutions to these equations.
	\end{quote} 
	\vfill
	\setcounter{footnote}{0}
\end{titlepage}
	
\clearpage
\setcounter{page}{2}
	
\tableofcontents
 
\section{Introduction}
\label{sec:intro}

M-theory, whose low energy limit is the 11-dimensional (11D) supergravity, is a framework that connects all 5 superstring theories under compactifications and dualities, and membranes and 5-branes are its solitonic
objects \cite{Hull:1994ys,Witten:1995ex,Townsend:1996xj}.
For a supersymmetric membrane of M-theory, as for a supersymmetric string of string theory, one has classically equivalent options for writing an action \cite{Duff:1996zn}:
\begin{enumerate}
    \item spinning membrane/string --- with manifest world-volume supersymmetry \cite{Howe:1977hp,Duff:1996zn,PhysRevD.3.2415,NEVEU197186,FRIEDAN198693,Deser1976ACA,Brink:1976sc}, 
    \item supermembrane/superstring --- with manifest target spacetime supersymmetry \cite{Bergshoeff:1987cm,Bergshoeff:1987qx,Green:1983wt,Green:1987sp},
    \item double supersymmetric membrane/string --- with manifest spacetime and world-volume supersymmetry \cite{Bandos:1995zw,Sorokin:1999jx}.
\end{enumerate}
The first two approaches are the most common ones in string theory and are known as Ramond-Neveu-Schwarz (RNS) and Green-Schwarz (GS) formalism respectively.
The RNS spinning string is formulated in terms of  a Polyakov type action \cite{PhysRevD.3.2415,NEVEU197186,FRIEDAN198693,Deser1976ACA,Brink:1976sc}, where one introduces an auxiliary two-dimensional world-volume metric, which carries three additional unphysical degrees of freedom. At the classical level they are compensated by the Weyl and diffeomorphism symmetries of the action, which also must be preserved at the quantum level. Vanishing of Weyl anomaly then imposes two sets of conditions. The first one requires the target space of the theory to be 10-dimensional (10D) whereas the latter sets the beta functions of the background fields to zero. At one loop level, which dominates the high string tension limit, this is equivalent to the 10D supergravity equations \cite{Callan:1989nz,CALLAN1987525,CALLAN1985593,Callan:1986bc,Callan:1985ia}.

Alternatively, the GS superstring is described by a superspace generalization of the Nambu-Goto action \cite{Green:1983wt,Green:1987sp}. In this formalism the string propogates in a supersymmetric target space with 10 bosonic and 32 anticommuting coordinates with supervielbein and super Kalb-Ramond fields set as a background. The restriction on the dimension of the target space again comes from the requirement of Weyl anomaly cancellation. On-shell the GS superstring has 8 bosonic and 16 fermionic degrees of freedom, 8 of which have to be eliminated to ensure supersymmetry, which is done by imposing $\kappa$-symmetry, whose job is to gauge away half of the propagating fermions. The requirement of $\kappa$-invariance  together with Bianchi identities for the supertorsion and the supertension condition for the super Kalb-Ramond field impose constraints that can be formulated in terms of a set of equations for background fields. In \cite{Wulff:2016tju} it has been shown that these equations are a generalization of the usual 10D supergravity which are currently referred as the {\it  generalized 10D supergravity}. The generalization is encoded in the appearance of an additional vector $I^m$ in the field equations, which is a Killing vector for the background fields. 
Explicitly equations of the 10D generalized supergravity are:
\begin{equation}
    \begin{aligned}    
        R_{mn} - \fr14 H_{mpq}H_n{}^{pq}  +2 \nabla_{(m}Z_{n)} & = T_{mn} ,\\
        -\fr12 \nabla^k H_{kmn} +Z^k H_{kmn} + 2\nabla_{[m}I_{n]} & = K_{mn},\\ 
        R - \fr12 |H_3|^2 + 4 (\nabla^m Z_m - I^mI_m-Z^mZ_m) & = 0,\\
        d *F_p - H_3 \wedge * F_{p+2} - \iota_I B_2\wedge *F_p -\iota_I *F_{p-2} &= 0,
    \end{aligned}
    \label{eq:gensugra0}
\end{equation}
where
\begin{equation}
    \begin{aligned}    
        T_{mn} & =\fr14 e^{2\Phi}\sum_p\left[\fr1{p!}F_m{}^{k_1\dots k_{p}}F_{nk_1\dots k_{p}} - \fr12 g_{mn}|F_{p+1}|^2\right], \\
        K_{mn}   &= \fr14 e^{2\Phi}\sum_p\fr1{p!}F_{k_1\dots k_p}F_{mn}{}^{k_1\dots k_p}.
    \end{aligned}
\end{equation}
We denote $|\w_p|^2 = \frac{1}{p!} \w_{i_1\dots i_p}\w^{i_1\dots i_p}$ for a $p$-form $\w_p$ and $Z_m=\dt_m\Phi + I^n B_{nm}$. Here $I=I^m\dt_m$ is a Killing vector field that is a symmetry of all the fields including the dilaton $\Phi$ and when it is zero we get the usual 10D supergravity equations. 
From the point of view of the RNS spinning string the requirement for the background fields to satisfy equations of generalized supergravity rather than the ordinary ones leads to breaking of Weyl symmetry to scale symmetry \cite{Arutyunov:2015mqj}. Recently there has been attempts to restore Weyl symmetry by generalizing the corresponding Fradkin-Tseytlin counterterm, which however might suffer from non-locality (for more details see \cite{Fernandez-Melgarejo:2018wpg,Muck:2019pwj}).

The narrative above is quite different from the historical development of this subject which begins with the works on integrable $\sigma$-models and their deformations that preserve integrability such as: deformations of the $SU(2)$ principal chiral model \cite{Cherednik:1981df}, Yang-Baxter (YB) $\sigma$-models on group manifolds \cite{Klimcik:2002zj,Klimcik:2008eq}, q-deformations of integrable $\sigma$-models \cite{Delduc:2013fga}. Particularly interesting results in this respect are the integrability of the Metsaev-Tseytlin superstring on the AdS${}_5\times \SS^5$ background \cite{Bena:2003wd} and its integrable $\eta$-deformation \cite{Arutyunov:2015qva,Arutyunov:2015mqj}. The corresponding $\eta$-deformed AdS${}_5\times \SS^5$ background 
has been found to violate 10D supergravity equations, but to preserve $\kappa$-invariance. Moreover, it is formally T-dual to a solution of Type IIB supergravity equations, the so-called HT background \cite{Arutyunov:2015mqj,Hoare:2015wia,Hoare:2015gda}. All fields of the HT background respect isometry generated by the Killing vector $I^m$ except the dilaton. Finally, in \cite{Arutyunov:2015mqj} the 10D generalized supergravity equations satisfied for such backgrounds were proposed and then derived in \cite{Wulff:2016tju} from the $\kappa$-invariance constraints.

In comparison to the spinning string, the situation for the spinning membrane of M-theory is significantly different as there is no Weyl symmetry to compensate  the unphysical degrees of freedom coming from an auxiliary world-volume metric \cite{Howe:1977hp,Duff:1996zn}. Hence, the most appropriate option for its description is the GS supermembrane formalism \cite{Hughes:1986fa,Bergshoeff:1987cm}. As in the string case, for the consistency of this approach the action is required to be $\kappa$-symmetric that  imposes restrictions on the background fields in which the membrane evolves. It has been long known, that the equations of 11-dimensional supergravity for the background fields are sufficient for the membrane to be $\kappa$-symmetric \cite{Howe:1997he,Bergshoeff:1987cm,Bergshoeff:1987qx}.

However, it is natural to ask whether a more general solution of membrane $\kappa$-symmetry constraints is possible, in analogy to the case of the GS superstring. The common lore says that while the generalization of the equations on the string theory side is related to breaking of Weyl symmetry down to scale symmetry, the membrane does not have Weyl symmetry from the very beginning, and hence there seems to be nothing to break. More technically, both for the superstring and the supermembrane the $\k$-symmetry implies that the dimension $\fr12$ torsion component is expressed in terms of a spinor superfield $\chi_\a$. In 10D after imposing the $\kappa$-symmetry, one has the option to require this to be a spinor derivative of the dilaton
or not according to which one gets either the ordinary or generalized supergravity~\cite{Wulff:2016tju}. However, in 11D there is no dilaton and hence there is no such choice. 

All these difficulties might make one doubtful about the existence of a generalization of 11D supergravity equations. On the other hand, absence of a 11D parent of 10D generalized supergravity would be rather puzzling since one would expect it to be part of the M-theory landscape. To resolve this conundrum, in \cite{Bakhmatov:2022rjn} we proposed a new approach that is based on
the bi-vector (Yang-Baxter) and tri-vector deformations of supergravity solutions in terms of double (DFT) and exceptional field theory (ExFT) \cite{Bakhmatov:2017joy,Bakhmatov:2018apn,Borsato:2020bqo,Gubarev:2020ydf}. They provide a T/U-duality covariant approach to supergravity and are formulated on an extended space endowed with a section constraint. We always assume this to be solved by keeping only the ordinary geometric coordinates. 
For more detailed review of DFT see \cite{Aldazabal:2013sca,Berman:2013eva,Hohm:2013bwa} and of ExFT see \cite{Baguet:2015xha,Musaev:2019zcr,Berman:2020tqn}. These theories contain enough degrees of freedom for description of such deformations as they are given by a local $O(10,10)$ transformation for the bi-vector case and a local transformation in the corresponding exceptional group for the tri-vector case. In the former case the deformation is defined by a bi-vector $\beta^{m n} = r^{i j} k_{i}{}^{m} k_{j}{}^{n}$, where $k_i{}^m$ are Killing vectors of the initial background and $r^{ij}=r^{[ij]}$ is a constant matrix. The deformation generates a solution of the 10D generalized supergravity equations~\eqref{eq:gensugra0} 
if the $r$-matrix satisfies the classical YB equation~\cite{vanTongeren:2015uha,Hoare:2016hwh, Orlando:2016qqu,vanTongeren:2016eeb,Araujo:2017jkb,Bakhmatov:2017joy,Hong:2018tlp,Borsato:2018idb,Bakhmatov:2018bvp}
\begin{equation}
    \begin{aligned}
        r^{ i_1[j_1}r^{j_2|i_2|}f_{i_1 i_2}{}^{j_3]}&=0,
    \end{aligned}
    \label{cyb}
\end{equation}
where $f_{i_2 i_2}{}^{i_3}$ are the structure constants of the isometry algebra $[k_{i_1},k_{i_2}]=f_{i_1 i_2}{}^j k_j$ and $I^m$ is \cite{Bakhmatov:2018apn}
\begin{equation} \label{I}
    I^m = r^{i_1 i_2}f_{i_1 i_2}{}^j k_j{}^m.
\end{equation}
Deformations with nonvanishing $r^{i_1 i_2}f_{i_1 i_2}{}^j$ are called {\it non-unimodular} since such an r-matrix defines a non-unimodular subalgebra of the full isometry algebra \cite{Hoare:2016hwh,Orlando:2016qqu} 
and generalized supergravity solutions correspond to non-unimodular 
YB deformations. 

This approach extends naturally to backgrounds of 11D supergravity by generalizing Yang-Baxter deformations \cite{Bakhmatov:2020kul,Gubarev:2020ydf,Bakhmatov:2019dow}. In this case generalized Yang-Baxter deformation is a local $E_{d(d)}$ rotation with a tri-vector parameter $\Omega^{mnk} = \frac{1}{3!} \rho^{i_1 i_2 i_3}k_{i_1}{}^mk_{i_2}{}^n k_{i_3}{}^k$, where $\rho^{i_1 i_2 i_3} = \r^{[i_1 i_2 i_3]}$ is a set of constants and $k_i{}^m$ are the Killing vectors of the initial background. For the deformation to generate solutions of supergravity a generalization of the classical Yang-Baxter equation must be satisfied \cite{Bakhmatov:2019dow}
\begin{equation}\label{gYBsl5}
    6 \rho^{[i_2| i_7 j_1} \rho^{|i_3 i_4| j_2} f_{j_1 j_2}{}^{|i_5]} + {\rho}^{j_1 j_2 [i_2} {\rho}^{i_3 i_4 i_5]} {f}_{j_1 j_2}\,^{i_7} = 0,
\end{equation}
together with  a generalization of the unimodularity condition
\begin{equation}\label{unimodExFT}
{\rho}^{i_1 i_2 i_3} {f}_{i_2 i_3}\,^{i_4} = 0.
\end{equation}
As it was shown in \cite{Gubarev:2020ydf} these conditions are sufficient for the generalized fluxes of ExFT to be invariant under the generalized YB deformation and hence for a deformed background to be a solution of the ordinary 11D supergravity. 

Now, the key idea of the approach we briefly reported in \cite{Bakhmatov:2022rjn} and to be described in detail in the current paper is to allow non-unimodular tri-vector deformations. Under these, fluxes of ExFT transform non-trivially acquiring additional terms containing the tensor
\begin{equation}
J^{m n} = \frac{1}{4}{\rho}^{i_1 i_2 i_3} {f}_{i_2 i_3}\,^{i_4} {k}_{i_1}\,^{m} {k}_{i_4}\,^{n},
\label{j}
\end{equation}
which is the 11D analogue of $I^{m}$ \eqref{I}. Note that this tensor vanishes for a unimodular deformation \eqref{unimodExFT}.
Rewriting equations of ExFT in terms of such shifted fluxes and decomposing all expressions in terms of 11D fields (in the split form) we arrive at a generalization of 11D supergravity equations. This procedure guarantees that non-unimodular tri-vector deformations (with some restrictions) solve the newly obtained equations similar to non-unimodular YB deformed 10D backgrounds solving equations of generalized supergravity.

Our most important findings can be summarized as follows: i) the proper generalization of the Killing vector $I^m$ is the tensor $J^{m n}$ \eqref{j}; ii) the GL(11) symmetry is broken to GL(11-d)$\times$ GL(d), which might be the desired symmetry breaking; iii) the derived equations reproduce those of the generalized 10D supergravity (in the split form) once an isometric direction is singled out. For simplicity we work in a certain truncation of 11D supergravity, which restricts the background to be of the form $M_7\times M_4$, where $M_7$ is a constant curvature space-time, and the only fields are the metric $g_{\m\n}$ on $M_7$, the metric $h_{mn}$ on $M_4$ and the 3-form field $C_{mnk}$ on $M_4$. 
The resulting equations are
\begin{equation}
    \label{eq:gengensugra0}
    \begin{aligned}
    0= &\ {\mR}_{m n}[h_{(4)}]  - 7\, \tilde{\nabla}_{(m}Z_{n)} +T_{mn} \\
     &\  + 8 (1+V^2)\Big(J_{(mn)}J^k{}_k - 2 J_{m k}J^{k}{}_n\Big) + 4 V_{m}V_{n}\Big(J^{kl}J_{kl} - 2 J^{kl}J_{lk}\Big)\\
     &\ +4 V_kV_{l}\Big(4 J_m{}^kJ_n{}^l -J^k{}_{m}J^l{}_{n} - 2 J^{kl}J_{(mn)} \Big)  + 8 V_{k}V_{(m} \Big( 2J^l{}_{n)}J^k{}_l - J J_{n)}{}^k - J J^k{}_{n)} + J^{kl}J_{n)l}\Big),\\
    0=&\ \fr17 e^{2\f}\, \mc{R}[\bar{g}_{(7)}] + \frac{1}{6}\, (\nabla V)^2 +\tilde{\nabla}^m Z_m - 6 Z_m Z^m  -2 J^{mn}J_{mn} + \fr43 J_{mn}J^{nm},\\
     0 =&\ \tilde{\nabla}^m F_{mnkl} - 6 \Big(Z^m F_{mnkl} + 2 J^{pm}C_{m[nk}J_{l]p} - J^{pm}J_{p[n}C_{kl]m}\Big),
    \end{aligned}
\end{equation}
where $F_{mnkl} = 4 \dt_{[m}C_{nkl]}$, $V^m = \fr{1}{3!}\e^{mnkl}C_{nkl}$, $ T_{mn} = -\fr13 h_{mn}(\nabla V)$ and
\begin{equation}
    \begin{aligned}
        Z_m & = \dt_m \phi - \fr23 \e_{mnkl}J^{nk}V^l ,\\
        \tilde{\nabla}_m & = \nabla_m - \dt_m \phi.
    \end{aligned}
\end{equation}
Setting $J^{mn}$ to zero above one gets equations of a truncated version of 11D supergravity found in \cite{Bakhmatov:2020kul}.
Finally, the tensor $J^{mn}$ must satisfy the following set of conditions
\begin{equation}
    \label{eq:condJ0}
    \begin{aligned}
    0&=J^{m [n} J^{k l]},\\
        0&=\dt_m J^{k l} + J^{k n}\dt_{n}e_m{}^a e_a{}^l+ J^{n l}\dt_{n}e_m{}^a e_a{}^k + J^{n l}\d_m{}^k \dt_n \phi,\\
        0&=\nabla_m\big(e^{-\f}J^{[mn]}\big) ,\\
        0&=J^{mn}\dt_n \phi ,\\
        0&=\nabla_{[m}Z_{n]}  - \fr13 J^{kl}F_{mnkl},\\
        0&=\nabla_k\Big(e^{-\phi}J^{k[l}V^{p]}\Big) ,\\
        0&=\nabla_k(J^{(pl)}V^k) - \nabla_k(V^{(p}J^{l)k}).
    \end{aligned}
\end{equation}
Interestingly enough these conditions imply that $S^{mn}=J^{(mn)}$ is a Killing tensor \cite{Bakhmatov:2022rjn}. Although looking very restrictive, these actually allow for new solutions that are beyond the standard 11D supergravity as we will illustrate in section \ref{solutions}. Hence, the theory is not trivial. It is important to mention, that the membrane $\kappa$-invariance has not been checked for the proposed equations and remains an open question for  further investigation. Therefore, throughout the article, we prudently refer to the proposed equations as {\bf a} generalization of 11D supergravity.

The paper is organized as follows. In Section \ref{sectionDFT} we first illustrate our method by deriving generalized 10D supergravity equations of motion (EoM) from non-unimodular generalized YB deformation. For this we briefly remind all necessary DFT ingredients and its flux formulation. Next we obtain constraints on flux transformations from DFT Bianchi identities and show that they require $I^{m}$ to be a Killing vector of the deformed background. We complete this part by a comparison of our method of obtaining generalized 10D supergravity equations with others in the literature.
In Section \ref{sectionExFT} we repeat all these for the SL(5) ExFT that describes 11D supergravity in the 11=7+4 split. We will explain some details of SL(5) ExFT and give its flux formulation for a particular truncation that we use to keep calculations relatively simple. Using ExFT Bianchi identities we find constraints \eqref{eq:condJ0} on $J^{mn}$. Finally, we  derive equations \eqref{eq:gengensugra0} for the generalization of 11D supergravity using flux shift procedure. Some explicit solutions of these equations, which illustrate richness of the theory are  presented in \ref{solutions}. Conclusions and further discussion are in Section \ref{sectionCD}. Some technical steps are explained in two appendices.

Large portion of our calculations were done using the computer algebra system Cadabra \cite{Peeters:2007wn} and the corresponding worksheets can be found at \cite{gengensugra_2022}.

\section{Generalized supergravity in 10 dimensions from DFT}\label{sectionDFT}

In this section we will show how equations of the 10D generalized supergravity can be derived from non-unimodular bi-vector Yang-Baxter deformations in DFT flux formulation. The advantage of this method is that it can easily be generalized to exceptional field theories and polyvector deformations.

\subsection{Flux formulation of DFT}

String theory possesses $O(d,d;\mathbb{Z})$ symmetry called T-duality on backgrounds of the form $\mathbb{T}^d \times \mM_{10-d}$ where $\mM_{10-d}$ is an appropriate $(10-d)$--dimensional spacetime  \cite{Duff:1989tf,Tseytlin:1990va,Tseytlin:1990nb,Siegel:1993th,Siegel:1993xq}. This symmetry is transmitted to the supergravity level as Cremmer-Julia $O(d,d;\mathbb{R})$ symmetry of supergravity equations (and the action for $d$ odd). Hence, Kaluza-Klein reduced supergravity on $\TT^d$  can be formulated in an explicitly T-duality covariant way and its solutions of the form $\mathbb{T}^d \times \mM_{10-d}$ fully preserve this symmetry. 

Let us now focus at the massless NS-NS closed string  sector ($g_{m n}, b_{m n}, \phi$) and discuss equations governing its low energy dynamics. The formalism that elevates the global $O(d,d;\RR)$ T-duality symmetry to the level of a local symmetry related to geometry of the underlying (extended) space is called double field theory. Initially this has been formulated as a field theoretical framework in \cite{Siegel:1993xq,Siegel:1993th} based on the earlier observation of doubling of coordinates of the closed string \cite{Fradkin:1984ai,Tseytlin:1990ar,Tseytlin:1990nb}. Its present name ``Double field theory'', its formulation in terms of generalized metric and further developments towards its geometric understanding, deriving large coordinate transformations and adding fermionic sector and supersymmetry has been done later in 
\cite{Hull:2009mi,Hull:2009zb,Hohm:2010jy,Hohm:2010pp,Hohm:2011dv,Geissbuhler:2013uka,Hull:2014mxa,Sazdovic:2017lqo}. For a comprehensive review of this approach see \cite{Aldazabal:2013sca,Berman:2013eva,Hohm:2013bwa}, while here we will focus only at formulation of DFT dynamics in terms of generalized fluxes. 

The bosonic sector of DFT consists of a generalized metric $\mH_{MN} \in O(10,10)/(O(10)\times O(10))$ and the invariant dilaton $d$, living in the doubled space $\XX^M = (x^m,\tx_m)$. Generalized Lie derivative acts on a generalized vector $V^M$ and $d$ as
\begin{equation}
    \label{eq:genlie}
    \begin{aligned}
         \mL_\L V^M =&\  \L^N \dt_N V^M - V^N \dt_N \L^M + \h^{MN}\h_{KL}\dt_N \L^K V^L,\\
         \mL_\L d  =&\ \L^M \dt_M d - \fr12 \dt_M \L^M \, .
    \end{aligned}
\end{equation}
Closure of these derivatives into an algebra of local transformations that is necessary for consistency of the theory requires the so-called section constraints 
\begin{equation}\label{2cond}
    \eta^{M N} \, \partial_{M} \, f \, \partial_{N} \, g = 0, \quad \eta^{M N} \, \partial_{M}  \, \partial_{N} \, f = 0,
\end{equation}
that are differential constraints for any fields $f$ and $g$ of the theory (first introduced in \cite{Siegel:1993th} and analyzed further in \cite{Berman:2012vc}) where 
\begin{equation}\label{invariantDFTmetric}
    \eta^{M N} = 
        \begin{pmatrix}
		       	0 & \delta^{m}{}_{n} \\
		    	\delta_{m}{}^{n} &  0
	    \end{pmatrix},
\end{equation}
which can be used to raise and lower indices.
More mathematically one may look at these constraints as conditions reducing the set of all functions of $\XX^M$ to a certain subset. Basically they project dependence of any fields to only 10 out of 20 $\mathbb{X}^{M}$ coordinates. In what follows we assume that nothing depends on $\tx_m$, while keeping all expressions formally containing $\tdt^m$ for covariance. In this case the formalism simply reproduces the standard supergravity, however, in more convenient variables. Keeping more general dependence on dual coordinates $\tx_m$ allows to describe non-geometric backgrounds and exotic branes (see \cite{Musaev:2019zcr,Berman:2020tqn} and references therein).

Equations of the theory can be derived from the following action
\begin{equation}\label{DFTactionQ}
    \begin{aligned}
    S_{DFT} = \int d^{10}\mathbb{X} \, e^{-2d} \,\Bigg[&
    \mH^{AB} \mF_A \mF_B + \mF_{ABC}\mF_{DEF}\left(\fr14\mH^{AD}\h^{BE}\h^{CF} - \fr{1}{12}\mH^{AD}\mH^{BE}\mH^{CF}\right)\\
    &- \mF_A \mF^A - \fr16 \mF_{ABC}\mF^{ABC}\Bigg],
    \end{aligned}
\end{equation}
where the so-called generalized fluxes of DFT are given by
\begin{equation}\label{DFTfluxes}
    \mF_{A B C} = 3 E_{N [C} \partial_{A} E^{N}{}_{B]},\qquad \mF_{A} = 2 \partial_{A} d - \partial_{M}{E^{M}{}_{A}}.
\end{equation}
Here $E^{N}{}_{A}$ is a generalized vielbein which transforms as a generalized vector (\ref{eq:genlie}) and $\mH^{A B}$ is the flat
$O(10,10)$ metric, $\mH^{M N} = E^{M}{}_{A} E^{N}{}_{B} \mH^{A B}$.
The flat derivative is defined as $\partial_{A} = E^{M}{}_{A} \partial_{M}$ where $\eta^{M N} = E^{M}{}_{A} E^{N}{}_{B} \eta^{A B}$:
\begin{equation}\label{DFTmetric}
    \eta^{A B} = 
        \begin{pmatrix}
		       	0 & \delta^{a}{}_{b} \\
		    	\delta_{a}{}^{b} &  0
	    \end{pmatrix},
	    \quad
    \mH^{A B} = 
        \begin{pmatrix}
		       	h^{a b} & 0 \\
		    	0 &  h_{a b}
	    \end{pmatrix}.
\end{equation}

To reproduce the NS-NS part of the 10D supergravity action with fields $e_{m}{}^{a}$, $b_{m n}$ and $\phi$ we use the so-called B-frame parametrization for the generalized vielbein and invariant dilaton
\begin{equation}
    \begin{aligned}
        d &= \phi - \frac12 \log\,e,\quad e = \det\, e_{k}{}^{a},\\
        E_M{}^{A} &= 
            \begin{bmatrix}
            e_m{}^{a} && 0 \\ \\
            - e^{k}{}_{a} b_{k m} && e^{m}{}_{a}
            \end{bmatrix}, \quad
            E^M{}_{A} = 
            \begin{bmatrix}
            e^m{}_{a} && - e^{k}{}_{a} b_{k m} \\ \\
            0 && e_m{}^{a}
            \end{bmatrix},
    \end{aligned}
\end{equation}
and substitute into \eqref{DFTactionQ}. Here we are interested in field equations rather than the action, for which we introduce variations 
\begin{equation}
\begin{aligned}
u^{B}{}_{A} = E_{M}{}^{B} \d E^{M}{}_{A} = \begin{bmatrix}
            e_{m}{}^{b} \d e^m{}_{a} && 0 \\ \\
            e^{p}{}_{a} e^{m}{}_{b} \d b_{m p} && - e_{m}{}^{a} \d e^m{}_{b}
            \end{bmatrix}, \quad
            \d d = \d \phi + \frac{1}{2} e_{m}{}^{a} \d e^m{}_{a},
\end{aligned}
\end{equation}
which explicitly keep the generalized vielbein in the B-frame. For more covariant approach to field variations of DFT based on projectors see \cite{Hohm:2010pp}. Then, variation of generalized fluxes becomes\footnote{See the files \texttt{F1\_variation}, \texttt{F2\_variation} and \texttt{udecomposition} of \cite{gengensugra_2022}.}
\begin{equation}
\begin{aligned}
\d \mF_{A} = \mF_{B} u^{B}{}_{A} - E^{M}{}_{B} \dt_{M} u^{B}{}_{A} + 2 E^{M}{}_{A} \dt_M \d d ,
\end{aligned}
\end{equation}
\begin{equation}
\begin{aligned}
\d \mF_{A B C} = 3 u^{D}{}_{[A} \mF_{B C] D} + 3 E^{M}{}_{[A} \eta_{C| D} \dt_{M} u^{D}{}_{|B]}.
\end{aligned}
\end{equation}
As a result, the full variation of the DFT action \eqref{DFTactionQ} and hence equations of the theory are written in terms of fluxes\footnote{See the file \texttt{EoM\_gen10} of \cite{gengensugra_2022}.}:
\begin{equation}\label{10fluxEOM}
\begin{aligned}
& e^{2d} \d \mL_{DFT} =  \d d \, (2\, {\mF}_{A} {\mF}_{B} {\mH}^{A B} - 4\, {\partial}_{B}{{\mF}_{A}}\,  {\mH}^{A B}   + \frac{1}{6}\, {\mF}_{A B C} {\mF}_{D\,  E F} {\mH}^{A D\, } {\mH}^{B E} {\mH}^{C F} \\
& - \frac{1}{2}\, {\mF}_{A B C} {\mF}^{A B}{}_{D}  {\mH}^{C D}- 2\, {\mF}_{A} {\mF}^{A}  + 4\, {\partial}_{A}{{\mF}^{A}}\, + \frac{1}{3}\, {\mF}_{A B C} {\mF}^{ABC}) \\
& + {u}^{A}\,_{B} \, \Big(2\, {\partial}_{M}{{\mF}_{C}}\,  {\mH}^{B C} {E}_{A}\,^{M} + \frac{1}{2}\, {\mF}_{A}{}^{ C D\, } {\mF}_{C D G}  {\mH}^{B G} + {\mF}_{A C D\, } {\mF}^{B C}{}_{ G} {\mH}^{D\,  G} + \frac{1}{2}\, {\mF}^{B}{}_{A\,  C} {\mF}_{D} {\mH}^{C D}\\
&  - \frac{1}{2}\, {\partial}_{M}{{\mF}^B{}_{A\,  C}}\ {\mH}^{C D} {E}_{D}\,^{M} + \frac{1}{2}\, {\mF}_{A C   D} {\mF}^{C} {\mH}^{B D} + \frac{1}{2}\, {\partial}_{D}{{\mF}_{A  E}{}^D}  {\mH}^{B E}  + \frac{1}{2}\, {\mF}^{B C}{}_{D\, } {\mF}^{D}  {\mH}_{C A}  \\
& + \frac{1}{2}\, {\partial}_{C}{{\mF}^{B C D}}\, {\mH}_{D A} - \frac{1}{2}\, {\mF}_{A C D\, } {\mF}_{E F G} {\mH}^{B E} {\mH}^{C F} {\mH}^{D\,  G} + \frac{1}{2}\, {\mF}_{C D}{}^{E} {\mF}_{F}  {\mH}^{B C} {\mH}^{D\,  F} {\mH}_{A E}    \\
&- \frac{1}{2}\, {\partial}_{G}{{\mF}_{C   E}{}^D}\,  {\mH}^{B C}{\mH}^{E G} {H}_{D\,  A} - 2\, {\partial}_{A}{{\mF}^{B}} - {\mF}_{A C D\, } {\mF}^{B C D} - {\mF}^B{}_{A C} {\mF}^{C} - {\partial}_{C}{{\mF}^{B C}{}_{A}}\Big).
\end{aligned}
\end{equation}
The first two lines above give the equation for the dilaton, which is precisely the same as in \cite{Geissbuhler:2013uka}. The terms with the $u^A{}_B$ prefactor contain the Einstein and the $B$-field equations. To see this let us decompose generalized fluxes in the $B$-frame, that give the following non-vanishing components
\begin{equation}
    \begin{aligned}
    \mF_{a b c} & = - H_{a b c},  \quad \mF_{a} = 2 e_{a}{}^{m} \nabla_{m} \phi + f_{a},\\
     \mF_{a b}{}^{c}& = f_{a b}{}^{c},\\
f_{ab}{}^{c} & = -2 e^{m}{}_{a} e^{n}{}_{b} \partial_{[m}e_{n]}{}^{c}, \qquad f_{a} = f_{ab}{}^{b}.
\end{aligned}
\end{equation}
The terms with the prefactors $\delta d$, $\delta e_a{}^m$ and $\delta b_{mn }$ give the standard supergravity equations 
\begin{eqnarray}\label{ordinary1}
\d d : && R - \frac{1}{12}\, H{}^{2} +  4\, {\nabla}^{m}{{\nabla_{m} \phi}}\, - 4\, (\nabla \phi)^2  = 0\,, \\ 
\d e^{m}{}_{a} : && R_{mn} - \frac{1}{4} H_{mkl} H_n{}^{kl} + \nabla_m \nabla_{n} \phi + \nabla_n \nabla_{m} \phi = 0\,, \\ \label{ordinary2}
\d b_{m n} : && \frac{1}{2}\, {\nabla}_{k}{{H}^{k m n}}\, - {H}^{k m n} \nabla_{k} \phi \,  = 0\,, \label{general3}
\end{eqnarray}
where $H_{m n k} = 3 \nabla_{[m} b_{n k]}$ and $\nabla_{m}$ is a covariant derivative with respect to metric $g_{m n} = e_{m}{}^{a} e_{n}{}^{b} g_{a b}$.

Let us emphasize that, in general terms contracting $\d e_m{}^a$ are not necessarily symmetric when the flat index is turned into a curved one. This is the case in the above calculation, where one ends up with the usual Einstein equation. However, this will no longer be true when we shift fluxes with a YB deformation in the upcoming sections. In this case the antisymmetric part of the equation  becomes proportional to a Lie derivative of the vielbein $e_m{}^a$ along the Killing vector $I^m$ and hence vanishes. 

\subsection{Non-unimodular Yang-Baxter deformation}

The flux formulation of equations of motion of the 10D supergravity that we reviewed above is very convenient for deriving their deformation. Indeed, on the one hand we know that solutions to equations of 10D generalized supergravity are generated from ordinary supergravity backgrounds by non-unimodular YB deformations. On the other hand, as we remind below, generalized fluxes of DFT do not transform under unimodular YB deformation but transform by a simple shift under their non-unimodular generalization. The narrative below is based mainly on  \cite{Bakhmatov:2017joy,Bakhmatov:2018apn,Borsato:2020bqo}. In terms of the generalized vielbein of double field theory a bi-vector deformation is simply an $O(10,10;\RR)$ rotation
\begin{equation}
    E'_{M}{}^{A} = O_{M}{}^{N} E_{N}{}^{A},
\label{rotation}
\end{equation}
that leaves the dilaton $d$ invariant. The rotation $O_M{}^N$ is defined as an exponent of the negative level generators of $O(10,10;\RR)$ w.r.t. the GL(10) decomposition:
\begin{equation}
            O_K{}^M = \exp\big(\b^{mn}T_{mn}\big) = \begin{bmatrix}
            \d^m{}_k && 0 \\
            \\
            \beta^{m k} && \d_m{}^k
            \end{bmatrix}.
\end{equation}
The generators of $O(10,10;\RR)$ satisfy
\begin{equation}
    [T_{MN},T_{KL}] = 2 \h_{K[M}T_{N]L} - 2 \h_{L[M}T_{N]K},
\end{equation}
where $T_{MN}$ are $20\times 20$ matrices. The bi-vector parameter $\b^{mn}$ is taken in the bi-Killing ansatz  $\beta^{m n} = r^{i j} k_{i}{}^{m} k_{j}{}^{n}$, where $\{k_{i}{}^{m}\}$ is a set of Killing vectors of the initial background and  $r^{i j}$ is a constant antisymmetric (Yang-Baxter) matrix. For the rotation
\eqref{rotation}
to act as a solution generating mechanism the $r$-matrix must satisfy 
\begin{equation}\label{10dcond}
    \begin{cases}
         f_{j_1 j_2}\,^{[i_1} r^{i_2|j_1|} r^{i_3]j_2} = 0, \qquad \text{(classical YB equation)},\\
         f_{i_1 i_2}\,^{j} r^{i_1 i_2} = 0, \qquad \text{(unimodularity)},
    \end{cases}
\end{equation}
which is also the condition for generalized fluxes to stay the same. Relaxing the second condition one obtains a non-unimodular YB deformation that amounts to the following transformation of fluxes\footnote{See the files \texttt{10D\_F1\_deformation} and \texttt{10D\_F3\_deformation} at \cite{gengensugra_2022}.}
\begin{equation}
    \label{eq:deffluxI}
    \d_{I} \mF_{A B C} = 0, \quad
    \d_{I} \mF_A = 2E_A{}^M
        \begin{bmatrix}
         0 \\ 
         I^{m}
        \end{bmatrix},
\end{equation}
where $I^m = \nabla_{k}\beta^{km} = f_{i_1 i_2}\,^{j} r^{i_1 i_2} k_{j}{}^{m} \neq 0$.

Here two important comments must be made. First, evidently after a YB deformation the generalized vielbein fails to remain in the $B$-frame and an additional $O(1,9)\times O(9,1)$ transformation is required. However, we do not have to find its specific form here as i) to recover the deformed background one simply uses generalized metric; ii) the trick to be described below explicitly uses generalized vielbein in the correct frame. The latter is related to our second comment, that is the deformation of fluxes  \eqref{eq:deffluxI} has the \emph{undeformed} vielbein on the RHS. The crucial observation is that this can be replaced by the deformed vielbein in all expressions. For exceptional field theory this will be true up to a slight constraint on the tri-vector deformation parameter.

\subsection{Constraints on \texorpdfstring{$I^{m}$}{I} from Bianchi identities}

To generate equations satisfied by the deformed fluxes \eqref{eq:deffluxI} let us consider a general shift of generalized fluxes 
\begin{equation}
    \label{eq:dftfluxX}
    \begin{aligned}
        \mF'_{ABC} &=\mF_{ABC}, \\
        \mF'_{A}-X'_{A} & = \mF_{A},
    \end{aligned}
\end{equation}
where fluxes on the RHS are constructed from a generalized vielbein $E_A{}^M$, the vector $X_A=E_A{}^MX_M = E'_A{}^M X_M = X'_A$ for some vielbein $E'_A{}^M $ and we define $X^M=(I^m,0)$. At this point we do not require anything from $I^m$ and hence the expressions $\mF'_{ABC}$ and $\mF'_{A}$ can not be interpreted as generalized fluxes in general. For that they must satisfy Bianchi identities, pretty much like the Bianchi identity for a Yang-Mills 2-form $dF=0$ allowing to write $F=dA$ as a field strength of a 1-form potential (at least locally). Hence, we require the LHS of \eqref{eq:dftfluxX} to satisfy Bianchi identities of DFT given that the RHS satisfies them \cite{Geissbuhler:2013uka}:
\begin{equation}
    \label{eq:BI_DFT}
    \begin{aligned}
        0&=\dt_{[A}\mF_{B C D]} - \fr34 \mF_{[AB}{}^E\mF_{CD]E},\\
        0&=2\dt_{[A}\mF_{B]}  + \dt^C \mF_{CAB}-\mF{}^C \mF_{CAB},\\
        0&=\dt{}^A \mF_A - \fr12\mF{}^A \mF_A + \fr1{12}\mF{}^{ABC}\mF_{ABC}.
    \end{aligned}
\end{equation}

Now, we would like to obtain field equations, that are identically satisfied when $E'_A{}^M = O^M{}_N E_A{}^N$ and $O^M{}_N$ is a non-unimodular Yang-Baxter deformation as defined above. Naturally, these must be equations of the generalized 10D supergravity. Given that, we can now show that acting on GL(10) scalars we have $\dt'_A =\dt_A $: 
\begin{equation}
    \begin{aligned}
        \dt'_A & = E'_A{}^m\dt_m = E_A{}^N O^m{}_N \dt_m\\
        &=E_A{}^m\dt_m + E_{A n}\b^{mn}\dt_m=E_A{}^M\dt_M = \dt_A,
    \end{aligned}
\end{equation}
where in the last line we used $\b^{mn}\dt_n=0$ due to the Killing vector conditions. Bianchi identities then give constraints on $X^M$. The first equation of \eqref{eq:BI_DFT} does not change, while the second becomes
\begin{equation}
    \begin{aligned}
    0&=2\dt_{[A}\mF_{B]}  + \dt^C \mF_{CAB}-\mF^C \mF_{CAB} =2\dt'_{[A}\mF_{B]}  + \dt'{}^C \mF_{CAB}-\mF^C \mF_{CAB}\\
    &=2\dt'_{[A}(\mF'_{B]} - X_{B]}) + \dt'{}^C \mF'_{CAB}-(\mF'{}^C-X^C) \mF'_{CAB} \\
    &= 2\dt'_{[A}\mF'_{B]} - 2 \dt'_{[A}X_{B]} + \dt'{}^C \mF'_{CAB}-\mF'{}^C \mF'_{CAB}+X^C \mF'_{CAB}\\
    &=  - 2 \dt'_{[A}X_{B]} +X^C \mF'_{CAB},
    \end{aligned}
\end{equation}
where in the last line we required of the deformed fluxes to satisfy Bianchi identities. Interestingly, the last line can be written as a generalized Killing vector condition for the new background $E'_M{}^A$:
\begin{equation}
    \begin{aligned}
       0&= 2 \dt'_{[A}X_{B]} -X^C \mF'_{CAB} \\
       &= 2 E'_{[A}{}^M\dt'_{M}\big( E'_{B]}{}^NX_{N}\big) 
       -X^M E'_M{}^C \big(2\dt'_{[A} E'_{B]}{}^N E'_{CN}+\dt'_{C} E'_A{}^N E'_{B N}\big)\\
       &=2 E'_{[A}{}^ME'_{B]}{}^N\dt_{M}X_{N} +2 E'_{[A}{}^M\dt_{M} E'_{B]}{}^NX_{N} -2 X_N  E'_{[A}{}^M \dt_M E'_{B]}{}^N 
       -X^M\dt_{M} E'_A{}^N E'_{B N}\\
        &=E'_{A}{}^ME'_{B}{}^N\dt_{M}X_{N}-E'_{B}{}^NE'_{A}{}^M\dt_{N}X_{M}-X^M\dt_{M} E'_A{}^N E'_{B N}\\
        &=-E'_{BN}\Big(X^M\dt_{M} E'_A{}^N
        -E'_{A}{}^M\dt_{M}X^{N}+E'_{A}{}^M\dt^{N}X_{M}\Big) = -E'_{BM} \mL_{X}{E'_{A}{}^M}.
    \end{aligned}
\end{equation}
Hence, the second Bianchi identity implies, that $X^M=E'_A{}^M X^A$ must be a generalized Killing vector of the \emph{deformed} generalized vielbein, i.e. of the metric and the $B$-field solving generalized supergravity equations. Note that we haven't used the relation between $\beta^{mn}$ and $I^m$.

Similarly the third Bianchi identity gives vanishing of the generalized Lie derivative of the invariant dilaton $d$ along $X^M$:
\begin{equation}
    \begin{aligned}
        0&=\dt'^A \mF_A - \fr12\mF^A \mF_A + \fr1{12}\mF^{ABC}\mF_{ABC}\\
        &=\dt'^A (\mF'_A-X_A) - \fr12(\mF'^A-X^A) (\mF'_A-X_A) + \fr1{12}\mF'^{ABC}\mF'_{ABC}\\
        &=-\dt'^A X_A + \mF'_AX^A - \fr12 X_A X^A\\
        &=-E'_A{}^M\dt_M(E'_N{}^AX^N) + (2 \dt_M d' - \dt_N E'_A{}^N E'_M{}^A) X^M - X_M X^M \\
        &= 2X^M \dt_M d' - \dt_M X^M - X_M X^M = 2\mL_X d' - X_M X^M.
    \end{aligned}
\end{equation}

Hence, we conclude that in order for the shifted generalized fluxes to again give generalized fluxes of a new vielbein and new invariant dilaton, these fields must have generalized isometry w.r.t. to the shift $X^M=(I^m,0)$. As we show below this is an exclusive property of double field theory: for exceptional field theory the vector $X^M$ is replaced by tensor $X_{MNK}{}^L$, that apparently does not have a similar nice geometric interpretation in terms of generalized isometries.

\subsection{10D generalized supergravity EoM}

We now turn to the derivation of equations of the 10D generalized supergravity from double field theory equations in flux formulation. The main idea is to start with equations written in terms of fluxes $(\mF_{ABC}, \mF_{A})$ and derivatives $\dt_A$ and use \eqref{eq:dftfluxX} to turn to new fluxes $(\mF'_{ABC}$, $\mF'_A)$. Since the initial fluxes correspond to a solution to DFT equations, we still get an identity now written in terms of $E'_M{}^A$ and $d'$, however with derivatives  $\dt_A= E_A{}^M\dt_M$. Finally, we recall, that one gets backgrounds of generalized supergravity from those of the ordinary supergravity by a non-unimodular YB deformation, that allows us to replace $\dt_A$ by $\dt'_A$. Hence, we arrive at a set of equations, that are by construction satisfied by backgrounds obtained as non-unimodular YB deformation of supergravity solutions, which must be precisely the desired generalized supergravity equations. Apparently, this procedure does not mean that all solutions to the new equations are related to ordinary supergravity backgrounds by such deformations. Indeed, after deriving the equations, one is free to search for solutions without referring to Yang-Baxter deformation techniques. 

\begin{figure}[ht!]
\begin{center}
\begin{tikzpicture}
    \node at (0,0) (E) {$E_M{}^A$};
    \node at (0,-2) (F) {$\mF_{ABC}$, $\mF_{A}$};
    \node at (0,-4) (Eoms) {EoMs($\mF_{ABC}$, $\mF_{A})=0$};
    \node at (0,-5) (Eomgr) {EoMs$_{SUGRA}(g,b,\phi) = 0$};
    

    \node at (6,0) (Ep) {$E'_M{}^A = (O_M{}^NE_N{}^A)|_{b-frame}$};
    \node at (6,-2) (Fp) {$\mF'_{ABC} = \mF_{ABC}$, $\mF'_{A} = \mF_{A} + X_{A}$};
    \node at (6,-4) (Eomsp) {EoMs($\mF'_{ABC}$, $\mF'_{A} - X_{A})=0$};
    \node at (6,-5) (Eomggr) {EoMs$_{I}(g',b',\phi') = 0$};    
    
    \draw [->, thick] (E)--(Ep);
    \draw [->, thick] (F)--(Fp);
    
    \draw [->] (F)--(Eoms);
    \draw [->] (Fp)--(Eomsp);    

    \draw [] (Eomgr)--(Eoms);
    \draw [] (Eomggr)--(Eomsp);
    
    \draw [<->, dashed] (E)--(F);
    \draw [<->, dashed] (Ep)--(Fp);
    
\end{tikzpicture}
	\caption{The algorithm to obtain generalized supergravity equations from DFT equations in the flux formulation by shifting generalized fluxes.}
	\label{figFluxApproach}
\end{center}
\end{figure}
Schematically the procedure is illustrated by Fig. \ref{figFluxApproach}, details of the pretty straightforward calculation can be found in the file \texttt{EoM\_gen10} at \cite{gengensugra_2022}. 

A comment must be made concerning the form of the generalized vielbein. As it has been mentioned above, the Yang-Baxter deformation spoil the upper-triangular form of the generalized vielbein by introducing a block proportional to $\beta^{mn}$. This however does not break the described procedure as we first determine how fluxes shift under non-unimodular deformation, then introduce a general shift $X^M$ of the flux. Only after imposing constraints on $X^M$ following from Bianchi identities we are able to define the generalized vielbein $E'{}_M{}^A$. Note, that this is written already in the upper triangular form and is related to the initial vielbein by a YB deformation plus an $O(1,9)\times O(9,1)$ rotation, whose particular form is not needed. The resulting equations are given in terms of the metric, $B$-field and the dilaton and the issue simply disappears.

\subsection{Comparison with other approaches}

Let us briefly comment on other approaches for deriving generalized 10D supergravity from extended field theories that are present in the literature. Here we do not follow the chronological order in favor of emphasizing possible relations with our approach.

\textbf{Modified DFT} approach of \cite{Sakatani:2016fvh} is probably the closest one to ours. Here the main idea is to construct a modification of DFT where all derivatives of the generalized dilaton $d$ are shifted as
\begin{equation}
    \nabla_M d \longrightarrow \nabla_M d + X_M. 
\end{equation}
This step looks very close to our shift of fluxes, if not the same, and was one of the inspirations in developing our approach. The above modification seems very natural since generalized supergravity backgrounds are formally T-dual along a coordinate $y$ to usual supergravity backgrounds with the dilaton linearly depending on $y$. All other fields are isometric w.r.t. $y$ and from the DFT point of view the formal T-duality  introduces a linear dependence on the dual coordinate $\tilde{y}$ in the dilaton \cite{Arutyunov:2015mqj,Catal-Ozer:2019tmm}. The additional terms coming from the above shift precisely corresponds to this dependence. However, in this form it is not clear how to generalize the shift to exceptional field theories and hence to 11D supergravity. We comment more on this possibility in the conclusion section \ref{sectionCD}.

\textbf{The exceptional field theory} approach of \cite{Baguet:2016prz} takes a similar perspective but focuses at dependence of backgrounds on dual coordinates. Equations of generalized supergravity are reproduced from equations of exceptional field theory by introducing a special Scherk-Shwarz ansatz with a twist matrix depending on a dual coordinate. The relationship of this method to ours is not very clear yet, however one would expect that the twist matrix produces precisely the required shift if the theory is written in terms of fluxes. This is an interesting direction for a more detailed investigation.

\textbf{The massive IIA from ExFT} derivation of \cite{Ciceri:2016dmd} does not reproduce generalized supergravity, however is ideologically very similar to what we describe here. The idea is to shift generalized Lie derivative as
\begin{equation}
    \mL_\L \longrightarrow \mL_\L + \L^M X_M,
\end{equation}
where $X_M$ is $\mathfrak{e}_{n}$ algebra valued and acts on tensors as a matrix $(X_M)_{K}{}^L$. Closure of the deformed Lie derivative imposes constraints on $X_M$, that in the original work is kept constant, and actually contains the Romans mass parameter. Although such a shift looks equivalent to our approach, given $X_M$ becomes a function, we were not able to make this relation precise. If possible, this would provide a better understanding of the flux shift procedure. 

\section{Generalization of eleven-dimensional supergravity}\label{sectionExFT}

We now would like to apply the above procedure to 11D supergravity equations formulated in terms of generalized fluxes of exceptional field theory. For simplicity we consider here the SL(5) ExFT, that corresponds to the 7+4 split of the 11D spacetime. Moreover, we will restrict ourselves to a truncated theory, where ExFT contains only scalar fields left after dimensional reduction to $D=7$. The whole procedure in principle can be repeated and extended to the full theory with any U-duality group, with possible subtleties in the E$_8$ case related to the extra gauge fields. For more detailed description of exceptional field theories the reader may refer to the original papers \cite{Berman:2010is,Berman:2011jh,Berman:2011pe,Hohm:2013jma,Hohm:2013pua,Hohm:2013vpa,Musaev:2014lna,Abzalov:2015ega,Musaev:2015ces} as well as to the reviews \cite{Baguet:2015xha,Musaev:2019zcr,Berman:2020tqn}. Here we are working with the SL(5) exceptional field theory whose scalar sector has been constructed in \cite{Berman:2010is} and the full action has been presented in \cite{Musaev:2015ces}.

\subsection{Basics of the SL(5) exceptional field theory}

Let us start with a brief reminder of the structure of the full exceptional field theory with the SL(5) group and its truncated version. Focusing only on the bosonic sector of 11D supergravity one finds that under the $11=7+4$ split the metric and the 3-form give rise to fields transforming as 7-dimensional tensors and collected into the following irreps of the SL(5) duality group:
\begin{equation}
    \begin{aligned}
        & \{g_{\m\n}, &&  A_{\m}{}^{MN}, && M_{MN}, && B_{\m\n M}\}.
    \end{aligned}
\end{equation}
Here capital Latin indices $M,N,\dots = 1,\dots,5$ label the $\bf 5$ of $\sl(5)$,  vector fields $A_\m{}^{MN} = - A_{\m}{}^{NM}$ belong to the $\bf 10$, $\mu, \nu,\dots = 0,\dots,6$, the generalized metric $M_{MN}$ labels the coset $\SL(5)/\USp(4)$. The theory is formally defined on a (7+10)-dimensional spacetime parametrized by coordinates $(x^\m,\XX^{[MN]})$ with additional constraint
\begin{equation}
    \e^{MNKLP}\dt_{MN}\bullet \otimes \, \dt_{KL}\bullet=0,
\end{equation}
usually referred to as the section condition \cite{Berman:2012vc}. Here bullets stand for any combination of fields and of their derivatives. In what follows we will always assume that the section condition is solved by keeping only dependence on 4 coordinates $x^m = \XX^{5m}$. This choice restores the standard supergravity (for further discussion in the framework of  exceptional generalized geometry see \cite{Coimbra:2011ky}). 
The section condition is necessary for local diffeomorphisms acting on tensors on the extended space to form a closed algebra. On a generalized vector of weight $\l$ these act as
\begin{equation}
    \d_\L V^M = \mc{L}_\L V^M = \frac{1}{2} \L^{KL} \dt_{KL} V^{M} - V^{L} \dt_{LK} \L^{M K} + \bigg(\frac{1}{4} + \lambda \bigg) V^{M} \dt_{KL} \L^{KL}.
\end{equation}
The Lagrangian of the full SL(5) theory is invariant under local symmetries including the above generalized Lie derivative as well as diffeomorphisms in the external 7-dimensional space, and reads
\begin{equation}\label{top}
    \begin{aligned}
     e^{-1}\mL =& \ \hat{R}[g_{(7)}] \mp \fr18 m_{MN}m_{KL}\mF_{\m\n}{}^{MK}\mF^{\m\n N L}+\frac{1}{48}g^{\m\n}\DD_{\m}m_{M N} \DD_\n m^{M N} + e^{-1}\mL_{sc}\\
     &+\fr{1}{3\cdot (16)^2}m^{MN}\mF_{\m\n\r M}\mF^{\m\n\r}{}_N+e\mL_{top}.
    \end{aligned}
\end{equation}
Covariant derivative $\mc{D}_\mu$ is defined in the standard Yang-Mills manner as follows
\begin{equation}
    \mc{D}_\m = \dt_\m - \mc{L}_{A_\m}.
\end{equation}
Its commutator $[\mc{D}_\m, \mc{D}_\nu]$ defines the field strength $\mc{F}_{\m\n}{}^{MN}$ and further tensor hierarchy, including the 2-form field $B_{\m\n M}$ and its field strength $\mF_{\m\n\r M}$.
The scalar part, that gives the scalar potential of maximal gauged supergravity upon generalized Scherk-Schwarz reduction (see e.g. \cite{Berman:2012uy} for the SL(5) theory), is given by
\begin{equation}
    \begin{aligned}
      e^{-1} \mL_{sc}=&\pm \Big(\frac{1}{8}\,  {\partial}_{MN}{{m}_{P Q}}\,  {\partial}_{KL}{{m}^{P Q}}\,  {m}^{M K} {m}^{N L} +\frac{1}{2}\, {\partial}_{MN}{{m}_{P Q}}\,  {\partial}_{KL}{{m}^{M P}}\,  {m}^{N K} {m}^{L Q} \\
     &+ \frac{1}{2}\, {\partial}_{MN}{{m}^{L N}}\,  {\partial}_{KL}{{m}^{M K}}+\fr12 m^{MK}\dt_{MN}m^{NL}(g^{-1}\dt_{KL}g)+\fr18 m^{MK}m^{NL}(g^{-1}\dt_{MN}g)(g^{-1}\dt_{KL}g)\\
     &+\fr18 m^{MK}m^{NL}\dt_{MN}g^{\m\n}\dt_{KL}g_{\m\n} \Big).
    \end{aligned}
\end{equation}
The topological term $\mc{L}_{top}$ in \eqref{top} is of no relevance here as will be set to zero by further truncation. The upper and lower signs corresponds to Minkowski and Euclidean signature of the metric on the external 7-dimensional space respectively. The generalized metric $m_{MN}$ is a generalized tensor of weight zero, the external metric $g_{\m\n}$ is a generalized scalar of weight $2/5$. 

Explicit parametrization of the generalized metric in terms of GL(4) fields reads
\begin{equation}\label{mmetric1}
\begin{aligned}
			m_{MN}=h^{\fr{1}{10}}
				\begin{bmatrix}
				h^{-\fr12}h_{mn} && -V_{m} \\ \\
				- V_{n} &&  \pm h^{\fr12}(1 \pm V_{k}V^{k})
			\end{bmatrix}, \quad 
	        m^{MN}=h^{-\fr{1}{10}}
				\begin{bmatrix}
				h^{\fr12}(h^{mn}\pm V^mV^n) && \pm V^{m} \\ \\
				\pm V_{n} &&  \pm h^{-\fr12}
			\end{bmatrix}			
\end{aligned}
\end{equation}
with $V^m=1/3! \, \ve^{mnkl}C_{nkl}$ and $h=\det h_{mn}$.	The fields $g_{\m\n}$ and $h_{mn}$ encoding metrics of the external and internal spaces of exceptional field theory respectively are related to the full 11D metric by the usual Kaluza-Klein ansatz
\begin{equation}
\label{KKgauge0}
  E_{\hat{\mu}}{}^{\hat{a}} \ = \ 
\begin{bmatrix} h^{-\fr1{5}}e_{\mu}{}^{\a} & &
  A_{\mu}{}^{m} h_{m}{}^{a} \\ \\ 0 & & h_{m}{}^{a}
  \end{bmatrix}\,,
\end{equation}
where $h = \det h_m{}^a$ denotes determinant of the vielbein $h_m{}^a$. 

Restricting dependence of all the fields to the coordinates $(x^\m, x^m = \XX^{5m})$ the above Lagrangian becomes precisely that of the 11D supergravity and the GL(11) symmetry gets restored. Hence, an important observation here is that the action either has the GL(11) symmetry or symmetry with respect to SL(5) duality transformations. As we will see below, the latter is necessary to obtain a generalization of supergravity equations of motion due to non-unimodular tri-vector generalized Yang-Baxter deformations. To determine the additional terms in the standard 11D supergravity equations we follow the framework of polyvector deformations developed in \cite{Bakhmatov:2020kul,Gubarev:2020ydf}, for which we turn to a truncated theory. It is worth mentioning here that, this step is not strictly necessary and is done only to simplify the computations. Extension to the full exceptional field theory is straightforward but rather tedious process. 

For the truncation we consider, background metric is in a block-diagonal form, i.e.\ $M_{11}=M_4\times M_7$, where the internal metric $h_{mn}$ does not depend on the external coordinates $x^\m$. In addition, equations of motion of the full SL(5) ExFT allow to assume consistently vanishing of the 1-form and 2-form fields, leaving us with the ansatz
\begin{equation}
\label{ansatz}
    \begin{aligned}
       &g_{\m\n}=g_{\m\n}(x^\m, x^m), && m_{MN}=m_{MN}(x^m), \\
       &A_{\m}{}^{MN}=0, && B_{\m\n\, M} =0.
    \end{aligned}
\end{equation}
By consistency we mean that the second line above does not impose further constraints on the remaining fields. Moreover, given the structure of the theory, the second line above can be applied already at the Lagrangian level. This simplifies the exceptional field theory setup, leaving us with only the $d=7$ Einstein-Hilbert term and the scalar potential for the generalized metric in the action. To obtain a theory defined completely in terms of a (rescaled) generalized metric we restrict dependence of the external metric on the coordinates $x^m$ as $g_{\m\n}(x^\m,x^m)=e^{\f(x^m)}h^{\fr15}\bar{g}_{\m\n}(x^\m)$. Finally, we rescale the external metric and the generalized metric as 
\begin{equation}\label{rescaling}
    \begin{aligned}
     h_{\m\n} & = e^{-2 \f}h^{\fr15}\bar{h}_{\m\n},\\
     m_{MN} & = e^{-\f}h^{\fr1{10}} M_{MN}.
    \end{aligned}
\end{equation}
that allows to rewrite the Lagrangian in the form similar to that of \cite{Berman:2010is, Blair:2014zba}
\begin{equation}
\label{L_cov_truncated}
    \begin{aligned}
       \mL=&\ \bar{e}\,M^{-1} \left(\mc{R}[\bar{h}_{(7)}]- \frac{1}{8}\, {M}^{K L} {M}^{M N} {\dt}_{K M}{{M}_{P Q}}\,  {\dt}_{L N}{{M}^{P Q}} - \frac{1}{2}\,  {\dt}_{N K}{{M}^{M N} }\,  {\dt}_{M L}{{M}^{K L}}\right. \\
       &+ \frac{1}{2}\, {M}^{K L} {M}^{M N} {\dt}_{M K}{{M}^{P Q}}\,  {\dt}_{P L}{{M}_{N Q}}  + {M}^{K L} {M}^{M N} {\dt}_{K P}{{M}_{M N}}\,  {\dt}_{L Q}{{M}^{P Q}} \\&-\left.\frac{15}{24}\, {M}^{K L} {M}^{M N} {M}^{P Q} {M}^{R S} {\dt}_{M P}{{M}_{K L}}\,  {\dt}_{N Q}{{M}_{R S}}  \right),
    \end{aligned}
\end{equation}
where $M=\det M_{MN}= e^{5\f}h^{-1/2}$, $\bar e = (\det \bar h_{\m\n})^{1/2}$ and $\mc{R}[h_{(7)}]$ is the Ricci curvature scalar of the metric $h_{\m\n}$. Given the truncation ansatz \eqref{ansatz} and \eqref{rescaling} the rescaled external metric $\bar{h}_{\m\n}$ satisfies\footnote{For details see the file \texttt{External\_SL5} of \cite{gengensugra_2022}.}
\begin{equation}\label{EOMforextmetric}
    \begin{aligned}
     \mc{R}_{\m \n}[\bar{h}_{(7)}] - \frac{1}{7} \bar{h}_{\m\n} \mc{R}[\bar{h}_{(7)}] = 0.
    \end{aligned}
\end{equation}
Here $\mc{R}[\bar{h}_{(7)}]=$constant which is true for the cases of interest, namely AdS$_4$ and $\SS^4$. 

For the rescaling~\eqref{rescaling} the full $d=11$ vielbein can be written in the following nice form
\begin{equation}
E_{\hat{\mu}}{}^{\hat{a}} \ = \ 
  \left(\begin{array}{cc} e^{-\f}\bar{e}_{\mu}{}^{a} &
  A_{\mu}{}^{m} h_{m}{}^{\alpha} \\ 0 & h_{m}{}^{\alpha}
  \end{array}\right),
\end{equation}
while the generalized metric becomes
\begin{equation}\label{mmetric}
\begin{aligned}
			M_{MN}=e^{\f}
				\begin{bmatrix}
				|h|^{-\fr12}h_{mn} && -V_{n} \\ \\
				- V_{m} &&  \pm |h|^{\fr12}(1\pm V_{k}V^{k})
			\end{bmatrix},
			\;\;
			M^{MN}=e^{-\f}
				\begin{bmatrix}
				|h|^{\fr12}(h^{mn}\pm V^mV^n) && \pm V_{n} \\ \\
				\pm  V_{m} &&  \pm |h|^{-\fr12} 
			\end{bmatrix} ,	
\end{aligned}
\end{equation}
with $V^m= \frac{1}{3!}\, \ve^{mnkl}C_{nkl}$ and $h=\det h_{mn}$. Substituting this into \eqref{L_cov_truncated} and dropping dependence on all coordinates but $x^m$ one gets for the Lagrangian:
\begin{equation}\label{lagr}
    \bar{e}^{-1}h^{-\fr12} \mL =  e^{-5\f} \mR[\bar{h}_{(7)}] +e^{-7\f}\left(\mR[h_{(4)}]+42 h^{mn}\dt_m \f\dt_n \f\mp\fr12 \nabla_{m}V^m\nabla_nV^n\right).
\end{equation}
Note that when $\mR[\bar{h}_{(7)}]=0$, the covariant Lagrangian \eqref{L_cov_truncated} reproduces the SL(5)$\times \RR^+$ Lagrangian of \cite{Blair:2014zba} up to total derivative terms.

\subsection{Flux Lagrangian and equations}

In our approach, to derive a generalization of 11-dimensional supergravity equations we need generalized fluxes and their transformations under non-unimodular generalized Yang-Baxter deformation. Hence, we now aim at equations of the truncated SL(5) theory in flux formulation, where proper non-unimodular shift of fluxes will give the desired generalization. The most transparent and straightforward way of doing this is to first rewrite the Lagrangian in terms of fluxes and then vary it taking into account that variation of fluxes can again be expressed in terms of fluxes.

Let us start with the Lagrangian and remind that generalized fluxes are defined as a generalization of anholonomy coefficients:
\begin{equation}
    \label{eq:genflux}
    \begin{aligned}
    \mL_{E_{AB}} E_C{}^M &= \mF_{ABC}{}^D E_D{}^M,\\
    \mF_{ABC}{}^{D} & = \frac32 E_{N}{}^{D} \partial_{[A B} E^{N}{}_{C]} - E^{M}{}_{C} \partial_{M N} E^{N}{}_{[B} \delta^{D}{}_{A]} -  \frac12 E^{M}{}_{[B|} \partial_{M N} E^{N}{}_{|A]} \delta^{D}{}_{C} .
    \end{aligned}
\end{equation}
Here $E_{A}^M$ is the inverse of the generalized vielbein defined as usual as $M_{MN} = E_{M}{}^A E_N{}^B M_{AB}$ with $M_{AB}=\mbox{diag}[1,1,1,1\pm1]$ being the diagonal matrix with the sign depending on the signature of the internal space as before. Explicitly for the generalized vielbein $E_M{}^A$ and its inverse $E_A{}^M$ one has
\begin{equation}\label{sl5vielbeinsandO}
            E_M{}^{A} = e^{\fr{\f}{2}}
            \begin{bmatrix}
            e^{-1/2} e_m{}^{a} && e^{1/2} V^a \\ \\
            0 && e^{1/2}
            \end{bmatrix}, \,
            E^M{}_{A} = e^{-\fr{\f}{2}}
            \begin{bmatrix}
            e^{1/2} e^m{}_{a} && 0 \\ \\
            - e^{1/2} V^{m} && e^{-1/2}
            \end{bmatrix}.
\end{equation}
The vielbein $E_{AB}{}^{MN}=4E_{[A}^{\,\, M}E_{B]}^{\,\, N}$ is in the $\bf 10$ irrep of SL(5) and we define $\partial_{AB}=E_{AB}{}^{MN}\partial_{MN}$.
Explicit check as in \cite{Berman:2012uy} shows that $\mF_{ABC}{}^D$ contains only components in the $\bf 10$, $\bf 15$ and $\bf \bar{40}$, which are called $\q_{AB},Y_{AB}$ and $Z^{ABC}$ respectively\footnote{See the file \texttt{SL5\_fluxes} of \cite{gengensugra_2022}.}
\begin{equation}
    \begin{aligned}
        \mF_{ABC}{}^D &= \frac32 Z_{ABC}{}^D + 5 \theta_{[AB}\d_{C]}{}^D +  \d_{[A}{}^{D} Y_{B]C}. 
    \end{aligned}
\end{equation}
The irreducible components can be written in terms of the generalized vielbein as follows
\begin{equation}
    \begin{aligned}
    \q_{AB} &= \frac{1}{10} E^M{}_{[A} \dt_{MN}E^N{}_{B]}-\frac{1}{10}E^{-1}E^{MN}{}_{AB}\dt_{MN}E, \\
    Y_{AB}&= - E^M{}_{(A} \dt_{MN}E^N{}_{B)},\\
    Z_{ABC}{}^{D}& = E^{M}_{[A} E^{N}_{B|} E_{K}^{D} \partial_{M N}{E^{K}_{|C]}} + \frac13 \Big(2 E^{M}_{[A|} \partial_{M N}{E^{N}_{|B|}}  + E^{M}_{[A} E^{N}_{B|} E^{-1} \partial_{M N}{E}\Big) \delta_{|C]}^{D}.
    \end{aligned}
\end{equation}

It is important to comment here on the difference between generalized Scherk-Schwarz reduction of the SL(5) theory performed in \cite{Berman:2012uy}, that relates it to $D=7$ maximal gauged supergravity, and flux formulation. In the former the generalized fluxes $\mc{F}_{ABC}{}^D$ are required to be constants and have the meaning of gaugings. In the latter these are not constants and represent degrees of freedom that transform conveniently under tri-vector deformations. As it has been observed already in \cite{Grana:2012rr} for DFT, to rewrite \eqref{L_cov_truncated} in terms of $\q_{AB}, Y_{AB}$ and $Z^{ABC}$ one has to add terms that vanish upon the section constraint:
\begin{equation}
    \begin{aligned}
        m \Delta \mc{L} = 3\, {\delta}_{M1}\,^{N1}\,_{M2}\,^{N2}\,_{M3}\,^{N3}\,_{M4}\,^{N4}  {\partial}_{N1 N2}{{E}_{A}\,^{M1}}\,   {\partial}_{N3N4}{{E}_{B}\,^{M2}}\,  {E}_{C}\,^{M3} {E}_{D\, }\,^{M4} {m}^{A C} {m}^{B D\, }.
    \end{aligned}
\end{equation}
Up to total derivative terms this is equivalent to\footnote{See the file \texttt{SL5\_flux\_form} of \cite{gengensugra_2022}.}
\begin{equation}
    \begin{aligned}
m \mL =&  - \frac{700}{3}\, {\q}_{A B} {\q}_{C D\, } {m}^{A C} {m}^{B D\, } + {Y}_{A B} {Y}_{C D\, } {m}^{A C} {m}^{B D\, } - \frac{1}{2}\, {Y}_{A B} {Y}_{C D\, } {m}^{A B} {m}^{C D\, } \\
     &+ \frac{9}{4}\, {Z}_{A B C}\,^{D\, } {Z}_{D\,  E F}\,^{A} {m}^{B E} {m}^{C F} + \frac{3}{4}\, {Z}_{A A1 B}\,^{C} {Z}_{D\,  E F}\,^{G} {m}_{C G} {m}^{A D\, } {m}^{A1 E} {m}^{B F} .
    \end{aligned}
\end{equation}

To express variation with respect to $E^M{}_{A}$ in terms of fluxes we define $u^B{}_A$ as follows
\begin{equation}
\begin{aligned}
\d E^M{}_A E^B{}_M \equiv u^B{}_A \equiv-\d E^B{}_M E^M{}_A.
\end{aligned}
\end{equation}
Components of $u^A{}_B$ encode variation of the internal vielbein $e_m{}^a$, the $\XX^{MN}$-dependent part of the external vielbein $\phi$ and of the 3-form $C_{mnk}$:
\begin{equation}
\begin{aligned}
{u}^{5}\,_{5} &= \frac{1}{2}\, {de}_{a}\,^{m} {e}^{a}\,_{m} - \frac{1}{2}\, \delta \phi ,\\
{u}^{a}\,_{5} &= - {\d V }^{m} {e}^{a}\,_{m} + {V}^{m} {e}^{a}\,_{m} {e}^{b}\,_{n} {\d e}_{b}\,^{n} ,\\
{u}^{a}\,_{b} &= {\d e}_{b}\,^{m} {e}^{a}\,_{m} - \frac{1}{2}\, {\d e}_{c}\,^{m} {e}^{c}\,_{m} {\delta}^{a}\,_{b} - \frac{1}{2}\, \delta \phi\, {\delta}^{a}\,_{b}.
\end{aligned}
\end{equation}
Note that $u^A{}_B$ is upper-triangular by construction, that is to ensure the variation keeps the theory in the supergravity frame.

Then variation of the fluxes can be written as follows
\begin{equation}
\begin{aligned}
\delta \theta_{A B} &=  -\frac{1}{10}{\partial}_{C [A}{{u}^{C}\,_{B]}}\, + \frac{1}{10} {\partial}_{A B}{{u}^{C}\,_{C}}\,  - 2 {u}^{C}\,_{[A} {\q}_{B] C},\\
d Y_{A B} & = {\partial}_{C (A}{{u}^{C}\,_{B)}}\,  + 2{u}^{C}\,_{(A} Y_{B) C},\\
\d Z^{E F D} &=- \frac{1}{16}\, {\partial}_{A B}{{u}^{D}\,_{C}}\,  {\epsilon}^{E F A B C} - 2{Z}^{E F ,[A} {u}^{D]}\,_{A} - 2{u}^{[F}\,_{A}{Z}^{E] A, D} + \frac{1}{48}\, {\partial}_{A B}{{u}^{C}\,_{C}}\,  {\epsilon}^{D E F A B} \\
&+ \frac{1}{24}\, {\partial}_{A B}{{u}^{A}\,_{C}}\,  {\epsilon}^{D E F B C}.
\end{aligned}
\end{equation}
Equations of motion in the flux formulation are\footnote{See the file \texttt{EoM\_varE} for ordinary supergravity equations and \texttt{EoM\_varE\_ALL}  for the described generalization.}
\begin{equation}
\label{eq:exft_eqns}
\begin{aligned}
&\frac{14}{3}\, {\partial}_{A C}{{\q}_{D E}}\,  {m}^{B D} {m}^{C E} + \frac{14}{3}\, {\delta}_{A}\,^{B} {\q}_{C D} {\q}_{E F} {m}^{C E} {m}^{D F} + \frac{14}{3}\, {\delta}_{A}\,^{B} {\partial}_{C D}{{\q}_{E F}}\,  {m}^{C E} {m}^{D F}- 4\, {Y}_{C D} {\q}_{A E} {m}^{B C} {m}^{D E} \\
&+ 2\, {\delta}_{A}\,^{B} {Y}_{C D} {Y}_{E F} {m}^{C E} {m}^{D F}  + 4\, {Y}_{A C} {Y}_{D E} {m}^{B D} {m}^{C E} - 2\, {\partial}_{A C}{{Y}_{D E}}\,  {m}^{B D} {m}^{C E} - {\delta}_{A}\,^{B} {Y}_{C D} {Y}_{E F} {m}^{C D} {m}^{E F} \\
&+ 2\, {Y}_{C D} {\q}_{A E} {m}^{B E} {m}^{C D} 
- 64\, {Z}^{C D B} {Z}^{E F G} {m}_{A G} {m}_{C E} {m}_{D F}+ 128\, {\delta}_{A}\,^{B} {Z}^{C D E} {Z}^{F G H} {m}_{C F} {m}_{D G} {m}_{E H}\\
 & + {\partial}_{A C}{{Y}_{D E}}\,  {m}^{B C} {m}^{D E} + 8\, {\q}_{C D} {Z}^{E F G} {\epsilon}^{B C D H A1} {m}_{A G} {m}_{E H} {m}_{F A1}
+ 4\, {\partial}_{C D}{{Z}^{E F G}}\,  {\epsilon}^{B C D H A1} {m}_{A G} {m}_{E H} {m}_{F A1}  \\
& + 64\, {Z}^{B C D} {Z}^{E F G} {m}_{A E} {m}_{C D} {m}_{F G}- 128\, {Z}^{B c D} {Z}^{E F G} {m}_{A E} {m}_{C F} {m}_{D G} + 128\, {\delta}_{A}\,^{B} {Z}^{C D E} {Z}^{F G H} {m}_{C E} {m}_{D F} {m}_{G H} \\
&- 2\, {Y}_{A C} {Y}_{D E} {m}^{B C} {m}^{D E} - 8\, {\q}_{C D} {Z}^{E F G} {\epsilon}^{B C D H A1} {m}_{A H} {m}_{E G} {m}_{F A1} - 64\, {Z}^{C D B} {Z}^{E F G} {m}_{A C} {m}_{D E} {m}_{F G}  \\
&- 4\, {\partial}_{C D}{{Z}^{E F G}}\,  {\epsilon}^{B C D H A1} {m}_{A H} {m}_{E G} {m}_{F A1}- 64\, {Z}^{B C D} {Z}^{E F G} {m}_{A D} {m}_{C E} {m}_{F G} =0.
\end{aligned}
\end{equation}
Before moving forward let us make a few comments on the equations we obtained. First, note that although $u^B{}_A$ is fixed to be in upper triangular form one can still consider equations of the theory above without a projection or a contraction with $u^B{}_A$ involved. The reason is that the $(\mbox{Eqns})^5{}_a$ components of the equations above are identically zero. This will not be true in the generalized case, where these impose conditions on the tensor $J^{mn}$. 
Second, it is useful to look at the equations above for the undeformed case in the supergravity frame, where they become
\begin{equation}
\begin{aligned}
&12\, {\nabla}^{m}{{\nabla}_{m}{\phi}\, } - 42\, {\nabla}^{m}{\phi}\,  {\nabla}_{m}{\phi}\,  - \frac{1}{2}\, {\nabla}_{m}{{V}^{m}}\,  {\nabla}_{n}{{V}^{n}} +(  2\, {e}^{a m} {\partial}_{m}{{f}_{a}}- \, {f}^{a} {f}_{a} \,  - \frac{1}{4}\, {f}^{a b}\,_{c} {f}_{a b}\,^{c} - \frac{1}{2}\, {f}^{a}\,_{b}\,^{c} {f}_{a c}\,^{b}) \, =0 , \\
 &{e}_{b m} {e}^{c n} {\partial}_{n}{{f}^{a}\,_{c}\,^{b}}\,  + {e}^{b}\,_{m} {e}^{c n} {\partial}_{n}{{f}_{b c}\,^{a}}\,   - {e}_{b m} {f}^{a c b} {f}_{c} - 2\, {\partial}_{m}{{f}^{a}}\,  - {e}^{b}\,_{m} {f}^{a c}\,_{d} {f}_{c b}\,^{d}- \frac{1}{2}\, {e}_{b m} {f}^{c d a} {f}_{c d}\,^{b} + {e}^{b}\,_{m} {f}^{c}\,_{b}\,^{a} {f}_{c} \\
&  - {e}^{b}\,_{m} {f}^{a}\,_{b}\,^{c} {f}_{c} + {e}_{b}\,^{n} {e}^{c}\,_{m} {\partial}_{n}{{f}^{a}\,_{c}\,^{b}}\, %
 + {e}^{b}\,_{m} {f}^{a}\,_{c}\,^{d} {f}_{b d}\,^{c} + 14\, {e}^{a n} {\nabla}_{m}{\phi}\,  {\nabla}_{n}{\phi}- 14\, {e}^{a n} {\nabla}_{m}{{\nabla}_{n}{\phi}\, }\\
 &+{e}^{a}\,_{m} \Big( {\nabla}_{m}{{V}^{m}}\,  {\nabla}_{n}{{V}^{n}}+{V}^{n} {\nabla}_{n}{{\nabla}_{k}{{V}^{k}} }- 2 {\nabla}^{n}{\phi}\,  {\nabla}_{n}{\phi} \, + 14\, {\nabla}^{n}{{\nabla}_{n}{\phi}\, }\,  - 7\,  {V}^{n} {\nabla}_{n}{\phi}\,  {\nabla}_{k}{{V}^{k}}\,   \Big) =0 , \\
 & 7\, {\nabla}_{m}{\phi}\,  {\nabla}_{n}{{V}^{n}}\,  - {\nabla}_{m}{{\nabla}_{n}{{V}^{n}}\, }\,  = 0,
 \end{aligned}
\end{equation}
where the standard anholonomy coefficients are defined as usual as
(see Appendix \ref{appRicciGTR} for details)
\begin{equation}
f_{ab}{}^{c} = -2 e_{a}^{m} e_{b}^{n} \partial_{[m}e_{n]}^{c}, \qquad f_{a} = f_{ab}{}^{b}
\end{equation}
and flat indices are raised and lowered by the flat metric $h_{ab}$ (here we choose Euclidean signature). Taking into account  $\nabla_m V^m = \fr1{4!}\e^{mnkl}F_{mnkl}$ the above can be massaged into
\begin{equation}
\label{eoms_V}
    \begin{aligned}
    &\d\f: &&  \fr57e^{2\f}\, \mc{R}[\bar{g}_{(7)}]+  \mR[h_{(4)}] + 12\, {\nabla}_{m}{{\nabla}_{n}{\f}\, }\,  {h}^{m n} - 42\, {\nabla}_{m}{\f}\,  {\nabla}_{n}{\f}\,  {h}^{m n} + \frac{1}{2}\, (\nabla V)^2=0, \\
    & \d V^{m}: && {\partial}_{m}(\nabla V) - 7\, (\nabla V) {\dt}_{m}{\f}=0, \\
    & \d h^{mn}: &&  {\mR}_{m n}[h_{(4)}]  - 7\, {\partial}_{m}{\f}\,  {\partial}_{n}{\f}\,  + 7\, {\nabla}_{m}{\nabla}_{n}{\f}\\
    & && +h_{m n}\left(-\fr12 e^{2\f}\mc{R}[\bar{g}_{(7)}]- \frac{1}{2}\, \mR[h_{(4)}] +28\, {\partial}_{k}{\f}\,  {\partial}_{l}{\f}\,  {h}^{k l}- 7\, {\nabla}_{k}{\nabla}_{l}{\f}\,  {h}^{k l} + \frac{1}{4}\, (\nabla V){}^{2} \right)=0 ,
    \end{aligned}
\end{equation}
which is the set of equations of the usual 11D supergravity in the split form for our truncation \cite{Bakhmatov:2020kul}.

\subsection{Non-unimodular tri-vector deformations}

To generalize the procedure of Section \ref{sectionDFT} to equations of 11-dimensional supergravity in the form of equations on generalized fluxes of the SL(5) exceptional field theory one has to determine transformation of $\mF_{ABC}{}^D$ under non-unimodular tri-vector deformation. The non-unimodularity parameter $J^{mn}$ to be defined below is the 11-dimensional analogue of the vector $I^m$. As it has been discussed in detail in \cite{Bakhmatov:2020kul,Gubarev:2020ydf} tri-vector deformations are defined as the following SL(5) transformation
\begin{equation}
    E'_A{}^M = O^M{}_N E_A{}^N, \quad O_M{}^N = \begin{bmatrix}
            \d_m{}^n && 0 \\
            \\
            \fr1{3!}\e_{mpqr}\W^{pqr} && 1
            \end{bmatrix},
\end{equation}
where $\e^{mnkl}$ is the Levi-Civita symbol and  the tri-vector $\W^{m_1m_2m_3}= \fr1{3!}\r^{i_1i_2i_3}k_{i_1}{}^{m_1}k_{i_2}{}^{m_2}k_{i_3}{}^{m_3}$ is defined in terms of Killing vectors $k_{i}{}^{m}$ and an antisymmetric constant matrix $\r^{i_1i_2i_3}$. The indices $i_k=1,\dots,N$ enumerate Killing vectors. Under tri-vector deformations generalized flux transforms as\footnote{This is shown explicitly in \texttt{deltaF\_to\_X} of \cite{gengensugra_2022}.}
\begin{equation}\label{Fdeformation8}
\begin{aligned}
    \delta_{\rho} \mF_{ABC}{}^{D} = & \, {E}^{m}\,_{A} {E}^{n}\,_{B} {E}^{k}\,_{C} {E}_{l}\,^{D} {J}^{l p} {\epsilon}_{m n  k p} \\
    & - \frac{1}{2}\, e_{(4)}^{-2} {E}_{5}\,^{E} {E}^{m n k}\,_{[A B C]} {k}_{i7 p} {k}_{i2 n} {k}_{i3 m} {k}_{i4 k} {k}_{i5}{}^{p} \big(6 \rho^{[i2| i7 j1} \rho^{|i3 i4| j2} f_{j1 j2}{}^{|i5]} + {\rho}^{j1 j2 [i2} {\rho}^{i3 i4 i5]} {f}_{j1 j2}\,^{i7}\big) ,
\end{aligned}
\end{equation}
where the expression in the parentheses in the second line is precisely the generalized classical Yang-Baxter equation. The first line is the non-unimodular part of the transformation where we define
\begin{equation}\label{Jbar}
J^{m n} = \frac{1}{4} {k}_{i_1}\,^{m} {k}_{i_4}\,^{n} {\rho}^{i_1 i_2 i_3} {f}_{i_2 i_3}\,^{i_4} = S^{(m n)} + I^{[m n]}.
\end{equation}

Observe that while in the case of bi-vector deformation the non-unimodular part had an interpretation of a generalized Killing vector, now this is not the case already due to the index count. Indeed, given generalized YB equation holds the transformation can be written as
\begin{equation}
    \d_\r \mF_{ABC}{}^D = X_{ABC}{}^D.
\end{equation}
To eliminate dependence on the generalized frame fields the X-tensor is defined as
\begin{equation}
    \label{eq:X2J}
    X_{mnk}{}^l = \e_{mnkp}J^{lp}.
\end{equation}
Hence $X_{MNK}{}^L$ does not include the metric and the C-field and therefore is a complete analogue of $X_M$. More complicated structure of the non-unimodular part of deformation in the exceptional case  compared to DFT causes certain issues one should be careful about. The first one, namely the lack of a generalized symmetry interpretation, has already been mentioned. The second one is that the equality $X_{ABC}{}^D=X'_{ABC}{}^D$ which we will be using in the procedure does not hold in general. Instead one finds
\begin{equation}
    X'_{ABC}{}^D = X_{ABC}{}^D - E_A{}^M E_B{}^N E_C{}^K E_5{}^D W_l X_{MNK}{}^l.
\end{equation}
The last term apparently vanishes when the deformations are restricted to bi-vector only and the theory reproduces generalized Type IIA supergravity in 10D. More generally one may require 
\begin{equation}
    \label{eq:condWJ}
        W_{l}\e_{mnkp}J^{lp}=0,
\end{equation}
which is true for all our examples. In principle, this is a condition on the deformation tensor and seems to only restrict the way one generates solutions to our generalized equations by non-unimodular tri-vector deformations. Indeed, the resulting set of equations to be presented below by construction does not contain the deformation tensor $W_m$, while $J^{mn}$ is understood as a set of additional parameters subject to certain constraints. Then the condition \eqref{eq:condWJ}, if satisfied, guarantees that backgrounds obtained by a non-unimodular tri-vector deformation of a solution to equations of 11-dimensional supergravity satisfy the generalized set of equations.

\subsection{Conditions and equations}

Given the explicit form of a non-unimodular tri-vector transformation of the generalized flux consider now a general shift 
\begin{equation}
    \label{eq:shift}
    \mF_{ABC}{}^D = \mF'_{ABC}{}^D - X_{ABC}{}^D,
\end{equation}
where $X_{ABC}{}^D = E'_{A}{}^ME'_{B}{}^NE'_{C}{}^KE'_{L}{}^DX_{MNK}{}^L$ and the only non-vanishing component of the latter is $X_{mnk}{}^l = \e_{mnkp}J^{lp}$. Now $J^{lp}$ is an arbitrary tensor, $E_A{}^M$ solves equations of the usual 11D supergravity and $E'_{A}{}^M$ will encode fields, that enter the set of generalized 11-dimensional equations. From the calculations below it follows that the equations are satisfied if i) $E'_{A}{}^M$ are related to $E_{A}{}^M$ by a non-unimodular tri-vector deformation, ii) the vector $J^{mn}$ is defined by \eqref{Jbar} and iii) the condition \eqref{eq:condWJ} holds. In this case non-unimodular tri-vector deformation is a solution generating transformation. 

It is crucial that $\mF'_{ABC}{}^D$ has precisely the same form as in \eqref{eq:genflux}, which is required for it to satisfy generalized Bianchi identities. Formally, Bianchi identities in exceptional field theory follow from the condition that generalized fluxes defined as non-linear expressions in terms of the vielbein $E_M{}^A$ and its derivatives, transform linearly under generalized Lie derivative, i.e.
\begin{equation}
    \label{eq:cov_flux_cond}
    \delta_\L \mF_{ABC}{}^D = \fr12 \L^{MN}\dt_{MN}\mF_{ABC}{}^D.
\end{equation}
Apparently, variation of the flux under 
\begin{equation}
\begin{aligned}
    \delta_{\Lambda} E_{C}{}^{M} = & \frac12 \Lambda^{A B} \partial_{A B}{E_{C}{}^{M}} - E_{C}{}^{L} \partial_{L K}{\Lambda^{M K}}
    + \frac14 E_{C}{}^{M} \partial_{K L}{\Lambda^{K L}} ,
\end{aligned}
\end{equation}
is precisely of that form, if one simply substitutes the above into the flux definition \eqref{eq:genflux} which is basically a consistency check. In order to obtain a condition on fluxes one rewrites the vielbein variation back in terms of flux as follows
\begin{equation}
\begin{aligned}
    \delta_{\Lambda} E_{C}{}^{M} = & \mF_{A B C}{}^{E} E_{E}{}^{M} \Lambda^{A B} - E_{A}{}^{M} \partial_{C B}{\Lambda^{A B}} + \frac14 E_{C}{}^{M} \partial_{A B}{\Lambda^{A B}}.
\end{aligned}
\end{equation}
Now the condition \eqref{eq:cov_flux_cond} gives non-trivial constraints on generalized fluxes called Bianchi identities
\begin{equation}
\begin{aligned}
    \mZ_{D F, A B C}{}^{E} = &\ \frac12 \partial_{A B}{\mF_{D F C}{}^{E}} + \frac12 \partial_{B C}{\mF_{D F A}{}^{E}} - \frac12 \delta_{A}^{E} \partial_{C G}{\mF_{D F B}{}^{G}}\\
    & - \frac14 \delta_{C}^{E} \partial_{B G}{\mF_{D F A}{}^{G}} + \frac14 \delta_{C}^{E} \partial_{A G}{\mF_{D F B}{}^{G}} + \frac12 \delta_{B}^{E} \partial_{C G}{\mF_{D F A}{}^{G}} \\
    & - \frac12 \partial_{A C}{\mF_{D F B}{}^{E}} - \mF_{B G C}{}^{E} \mF_{D F A}{}^{G} + \mF_{A G C}{}^{E} \mF_{D F B}{}^{G} \\
    & + \mF_{A B G}{}^{E} \mF_{D F C}{}^{G} - \mF_{A B C}{}^{G} \mF_{D F G}{}^{E} - \frac12 \partial_{D F}{\mF_{A B C}{}^{E}} = 0.
\end{aligned}
\end{equation}
For clarification let us comment on the logic here. One can approach Bianchi identities from two different perspectives. The first one is to start with the explicit definition of flux as in \eqref{eq:genflux}, in which case Bianchi identities hold identically. The condition \eqref{eq:cov_flux_cond} is satisfied and the expression \eqref{eq:genflux} is proved to be a generalized scalar. The other perspective is to start with a tensor $\mF_{AB,C}{}^D$ that is required to satisfy Bianchi identities. In this case  \eqref{eq:genflux} appears as an explicit solution to Bianchi identities expressing $\mF_{ABC}{}^D$ in terms of 
generalized vielbein $E_M{}^A$. This is similar to the situation in Yang-Mills theory, where one (locally) solves Bianchi identities $dF=0$ as $F=dA$. Bianchi identities of the Yang-Mills theory can be obtained in a similar manner; for illustration consider the Abelian theory with $F_{\m\n} =2 \dt_{[\mu}A_{\n]}$. Now one requires this expression to transform as a tensor under coordinate shift parametrized by $\x^\mu$, which is apparently true. To derive a non-trivial condition we write
\begin{equation}
    \d_\x A_\m = \x^\nu \dt_\n A_\m + A_\n \dt_\m \x^\n = \x^\n F_{\n\m} + \dt_\m(A_\nu \x^\nu).
\end{equation}
For the field strength this transformation gives
\begin{equation}
    \d_\x F_{\m\n} = 2 \dt_{[\m}\d_\x A_{\nu]} = L_\x F_{\m\n} - 3 \x^\r \dt_{[\m}F_{\n\r]},
\end{equation}
where $L_\x$ denotes the standard Lie derivative and the last term is precisely the Bianchi identities.

We will stick to the understanding of Bianchi identities as a requirement for the generalized flux to be expressible in terms of vielbein. Hence, it is natural to require the shifted generalized flux $\mF'_{ABC}{}^D$ to satisfy the same Bianchi identities, so it can be expressed in terms of some (deformed) generalized vielbein $E'_M{}^A$. Substituting $\mF= \mF'-X$ into Bianchi identities and taking into account that these hold for $\mF_{ABC}{}^D$ we obtain (antisymmetrization in $[AB]$ and $[CD]$ is assumed)
\begin{equation}
    \begin{aligned}
    0&=\frac{1}{2}\, {\partial'}_{C D\, }{{X}_{A B E}\,^{F}}\,  - {\partial'}_{C E}{{X}_{A B D\, }\,^{F}}\,   - \frac{1}{2}\, {\partial'}_{A B}{{X}_{C D\,  E}\,^{F}}\,+ \frac{1}{2}\, {\delta}_{E}\,^{F} {\partial'}_{C G}{{X}_{A B D\, }\,^{G}}  - {\delta}_{C}\,^{F} {\partial'}_{E G}{{X}_{A B D\, }\,^{G}} \\
    & - 2\, {X}_{A B C}\,^{G} {\mF'}_{D\,  G E}\,^{F} + {\mF'}_{A B E}\,^{G} {X}_{C D\,  G}\,^{F} + {X}_{A B E}\,^{G} {\mF'}_{C D\,  G}\,^{F} - {\mF'}_{A B G}\,^{F} {X}_{C D\,  E}\,^{G}\\
    & - {X}_{A B G}\,^{F} {\mF'}_{C D\,  E}\,^{G}  + 2\, {\mF'}_{A B C}\,^{G} {X}_{D\,  E G}\,^{F}
      - X_{B G C}{}^{E} X_{D F A}{}^{G} + X_{A G C}{}^{E} X_{D F B}{}^{G}  + X_{A B G}{}^{E} X_{D F C}{}^{G} ,
    \end{aligned}
\end{equation}
where as before $\dt'_{AB} = E'_{A}{}^ME'_{B}{}^N\dt_{MN} = \dt_{AB}$ due to the deformation ansatz:
\begin{equation}
    \begin{aligned}
        \dt'_{AB}&= E'_{A}{}^ME'_B{}^N\dt_{MN} = E'_{A}{}^KE'_B{}^LO^M{}_K O^N{}_L\dt_{MN} =
        2E'_{[A}{}^KE'_{B]}{}^LO^{5}{}_K O^m{}_L\dt_{5m}\\
        &=2E'_{[A}{}^5E'_{B]}{}^lO^{5}{}_5 O^m{}_l\dt_{5m}+2E'_{[A}{}^kE'_{B]}{}^lO^{5}{}_k O^m{}_l\dt_{5m}\\
        &=E_{A}{}^ME_B{}^N\dt_{MN}+2E'_{[A}{}^kE'_{B]}{}^m W_k \dt_{5m}.
    \end{aligned}
\end{equation}
The first term is of the desired form while to show that the second term is vanishing on ExFT scalars we write
\begin{equation}
    \begin{aligned}
        W_{[m}\dt_{n]} = \e_{mnkl}\e^{klpq}W_p\dt_q = \e_{mnkl}\W^{klq}\dt_q \simeq 0.
    \end{aligned}
\end{equation}

In analogy to the case of double field theory we consider constraints coming from terms linear and quadratic in $X_{ABC}{}^{D}$ separately. For the latter we have
\begin{equation}
    \begin{aligned}
        - \frac{1}{16}\, {E}^{k}\,_{A} {E}^{l}\,_{B} {E}^{m}\,_{C} {E}^{n}\,_{D} {E}^{p}\,_{F} {E}_{q}\,^{E}  {J}^{q [r} {J}^{s t]}( {\epsilon}_{k l m n} {\epsilon}_{p r s t} - {\epsilon}_{k l m p} {\epsilon}_{n r s t}) = 0 ,
    \end{aligned}
\end{equation}
which is satisfied if
\begin{equation}
    J^{m [n} J^{k l]} = 0.
\end{equation}
Linear conditions require more work and eventually the full list of conditions takes the form \eqref{eq:condJ0}.
Details of the derivation of these conditions are collected in Appendix \ref{app:BI_der} and can be tracked in the Cadabra file \texttt{BI\_to\_X} of \cite{gengensugra_2022}. Interestingly, the fourth linear condition on $J$ ensures that equations following from the  antisymmetric part of $\delta e_m{}^a$ are satisfied identically. This is essential, as only the symmetric part gives a generalization of the Einstein equation. Similarly in the 10D case the antisymmetric part is zero given $I^m$ is a Killing vector. Multiplying the first linear condition on $J$ with $e_a{}^m$ we can rewrite it as
\begin{equation}
    L_{e_a}J^{kl} + J^{nl}\dt_n \phi e_a{}^k=0,
\end{equation}
where $L_{e_a}$ is the standard Lie derivative along $e_a = e_a{}^m\dt_m$. Equivalently this can be rewritten as
\begin{equation}
    [e_a, J]^{kl}+J^{nl}\dt_n \phi e_a{}^k=0,
\end{equation}
where the bracket $[\,,]$ is the Schouten-Nijenhuis bracket of degree $(1,2)$ defined as
\begin{equation}
    [A,B]^{m_1\dots m_{p+q-1}} = p A^{n[m_1\dots m_{p-1}} \dt_n B^{m_p\dots m_{p+q-1}]} + q(-1)^{pq} B^{n[m_1\dots m_{q-1}} \dt_n A^{m_q\dots m_{p+q-1}} ,
\end{equation}
for antisymmetric tensors and similarly for symmetric ones.


The final step for deriving equations of the generalization of 11D supergravity is to substitute flux shift as in \eqref{eq:shift}, with $X_{ABC}{}^D$ given by \eqref{eq:X2J} contracted with the proper generalized vielbein, to the exceptional field theory equations \eqref{eq:exft_eqns}. Taking into account conditions on $J^{mn}$ the equations can be written compactly as in \eqref{eq:gengensugra0}.\footnote{See the file EoM\_varE\_ALL of \cite{gengensugra_2022}.}

\subsection{Solution examples}\label{solutions}
Although conditions \eqref{eq:condJ0} on the tensor $J^{mn}$ seem to be quite severe, the theory is not empty and permits interesting solutions.
Below we present three examples, obtained by performing a non-unimodular tri-vector deformation on the AdS${}_4\times \SS^7$ background. The $J$-tensor for all these solutions satisfies conditions \eqref{eq:condJ0}. Note however, that although these conditions are essential for derivation of the equations \eqref{eq:gengensugra0} from tri-vector deformations, one may consider this set of equations without reference to deformations. In this case, conditions \eqref{eq:condJ0} can be ignored and then theory would allow solutions beyond non-unimodular tri-vector deformations. At least, at our current understanding we haven't found any other consistency constraints that would require \eqref{eq:condJ0}.

Let's arrange the deformations of $AdS_4 \times S^7$ by powers of coordinates inside the tri-vector expressed as $\W= \frac{1}{3!}\, \r^{abc}k_a \wedge k_b \wedge k_c$. We will use the Killing vectors of the $AdS_4$ space 
given as:
\begin{equation}
\begin{aligned}
&&P_{a} &= \dt_a, & K_a &= x^2 \dt_{a} + 2 x_a D,\\
&&D&=-x^m\dt_m, &  M_{ab} &= x_a \dt_b - x_b \dt_a,
\end{aligned}
\end{equation}
where $a,b=0,1,2$ and $m,n=0,1,2,z$, and we define $x^2 = \h_{mn} x^m x^n$ and $x_a = \h_{ab}x^b$.

\

$\bullet \, \mathbf{ P\wedge P \wedge M}$

\

We first consider the deformation with a linear tri-vector
\begin{equation}
    \W = \fr18 \r^{ab,cd} P_a\wedge P_b \wedge M_{cd} = \fr12 \left[ (\a-\a')x_0 + (\b-\b')x_1 + (\g-\g')x_2 \right] \partial_0 \wedge \partial_1 \wedge \partial_2,
\end{equation}
where
\begin{equation}
    \begin{aligned}
        \a &= \r^{01,02},\\ 
        \a' &= \r^{02,01},
    \end{aligned}
    \quad
    \begin{aligned}
        \b &= \r^{01,12},\\ 
        \b' &= \r^{12,01},
    \end{aligned}
    \quad
    \begin{aligned}
        \g &= \r^{02,12},\\ 
        \g' &= \r^{12,02} \, .
    \end{aligned}
\end{equation}
The resulting background was originally found in~\cite{Bakhmatov:2020kul} and although $\W$ is non-Abelian, it is a solution of the ordinary 11D supergravity {\it for any values of} $\r$: 
\begin{equation}
\begin{aligned}
ds^2 &= \fr{R^2}{4z^2} K^{-\fr23} \left[-(dx^0)^2+(dx^1)^2+(dx^2)^2 \right] + R^2 K^{\fr13} \left[ \fr{dz^2}{4 z^2} + d\W_{(7)}^2 \right],\\
F &= -\fr{3 R^3 }{8 z^4}  K^{-2}\, dx^0 \wedge dx^1 \wedge dx^2 \wedge dz,\\
K & = 1 + \fr{\r_a x^a}{z^3},
\end{aligned}
\end{equation}
where $\r_0 = \a-\a',\, \r_1 = \b-\b',\, \r_2 = \g-\g'$ and $R$ is the $AdS$ radius. 

In the present context we are interested in this deformation for a different reason: {\it a specific choice of} $\r$ renders it a solution to generalized 11D supergravity. Indeed, examination of the quadratic Bianchi constraints $J^{m[n} J^{kl]} = 0$ yields as one possible solution
\begin{equation}
\a = -\a',\quad \b = -\b',\quad \g = -\g'.
\end{equation}
With this choice, the generalized Yang-Baxter equation~\eqref{gYBsl5} reduces to a single constraint:
\begin{equation}
   \a^2 = \b^2 + \g^2.
\end{equation}
For any $\r$ satisfying this, the deformed background is non-unimodular and is a solution to generalized supergravity with $J = \fr14 \left( \a\, \partial_2 \wedge \partial_1 + \b\, \partial_2 \wedge \partial_0 + \g\, \partial_0 \wedge \partial_1 \right)$.

\

{$\bullet \, \mathbf{ D\wedge P\wedge P}$}

\

This deformation was first found in~\cite{Bakhmatov:2020kul}. The $\r$-matrix is three-parametric and the tri-vector is
\begin{equation}
    \W = \fr{2}{R^3}\,D\wedge (\r_a \e^{abc} P_b \wedge P_c).
\end{equation}
The $\r$-tensor chosen in this way satisfies the generalized Yang-Baxter equation for any values of $\r_a$. The solution to the generalized 11-dimensional supergravity equations reads
\begin{equation}
\begin{aligned}
ds^2 &= \fr{R^2}{4z^2} K^{-\fr23} \left[-(dx^0)^2+(dx^1)^2+(dx^2)^2 +\left( 1+ \frac{\r_a x^a}{z^3} \right)\,dz^2 -\fr1{z^2}\r_a dx^a dz \right]+ R^2K^{\fr13} d\W_{(7)}^2,\\
F &= -\fr{3 R^3 }{8 z^4}  \left( 1+ \frac{\r^2}{12 z^4} \right) K^{-2}\, dx^0 \wedge dx^1 \wedge dx^2 \wedge dz,\\
J^{ab} &= - \fr{4}{3R^3}\e^{abc}\r_{c}, \\
K & = 1 + \fr{\r_a x^a}{z^3} - \frac{\r^2}{4z^4} ,
\end{aligned}
\end{equation}
where we denote $\r^2=\r_a\r_b\h^{ab}$. It is worth mentioning, that for $\r^2 = 0$ the deformed background solves equations of motion of the usual supergravity~\cite{Bakhmatov:2020kul}. This is an 11-dimensional example of what was called trivial solutions to generalized supergravity equations in~\cite{Wulff:2018aku}, when terms with $I^m$ (or in the present case $J^{mn}$) vanish separately.

\

{$\bullet \, \mathbf{ D\wedge M\wedge M}$}

\

Our last example is a non-trivial solution to the equations of the 11D generalized supergravity and is based on the deformation given by the tri-vector
\begin{equation}
    \W = \frac{4}{R^3}\, \r_a \e^{abc}\, D\wedge M_{bd} \wedge M_c{}^d = \frac{4}{R^3}\, \r_a x^a\, \left( x^b x_b \, \dt_0 \wedge \dt_1 \wedge \dt_2 - \fr12 z\,  \e^{bcd} x_b\, \dt_c \wedge \dt_d\wedge \dt_z\right).
\end{equation}
The $\r$-matrix should have  $\r^2 = \r_a \r_b \h^{ab} = 0$ due to the generalized YB equation
and the solution is
\begin{equation}
\begin{aligned}
ds^2 =&\ \frac{R^2}{4 z^2}\, K^{-\fr23} \left[ dx_a dx^a + \frac{1}{z^2}\, \r_a x^a x^b dx_b dz+\left( 1 - \frac{x_a x^a \r_b x^b}{z^3} \right) dz^2 \right]+ R^2 K^{\fr13} d\W_{(7)}^2,\\
F_{012z} =&\ -\fr38  \frac{R^3}{z^4} K^{-2} \left( 1 + \frac{1}{12} \frac{x_a x^a \r_b\r_c x^b x^c}{z^4} \right),\\
J^{ma} &= \fr{32}{R^3}\r_b\e^{abc}x_cx^m, \\
K & = 1 + \frac{x_a x^a}{z^3}\, \r_b x^b \, \left( 1 - \frac{\r_c x^c}{4z} \right).
\end{aligned}
\end{equation}

\section{Conclusions and discussions}\label{sectionCD}

In this paper we present a detailed description of the generalization of equations of 11D supergravity announced in \cite{Bakhmatov:2022rjn}. These equations are satisfied by non-unimodular tri-vector generalized YB deformed supergravity backgrounds. In this respect our result is a natural extension 10D generalized supergravity to 11D. The approach we develop here is based on the formalism of exceptional and doubled field theory, where (generalized) YB deformations become simply a local T/U-duality transformation, that preserve fluxes. Non-unimodularity then shifts fluxes generating the Killing vector $I^\mu$ in 10D and a tensor $J^{mn}$ with no index symmetry in 11D. The tensor $J^{mn}$ enters explicitly into the generalized equations \eqref{eq:gengensugra0} and additionally satisfies a list of conditions \eqref{eq:condJ0}. Although looking quite restrictive, these allow non-trivial solutions several examples of which are given here. 

The equations obtained contain those of 10D generalized supergravity for particular backgrounds of the form $M_7\times M_3$, that is dictated by the formalism and the truncation we used. In general, we expect no obstacles for repeating the derivation for arbitrary backgrounds, which is however an interesting task for the reason we discuss below. The reduction to 10D goes basically along the lines of the reduction of the scalar sector of SL(5) ExFT to the $O(3,3)$ DFT \cite{Thompson:2011uw}. On top of that one must restrict to only bi-vector deformations, which is done by choosing on of the Killing vectors entering the deformation ansatz to commute with the others. Choosing the adapted basis and denoting the corresponding direction $x^*$ we decompose the tensor $J^{mn}$ as
\begin{equation}
    \begin{aligned}
        J^{mn} && \to && J^{**}, && J^{*a}, && J^{a*}, && J^{ab} ,
    \end{aligned}
\end{equation}
where $m,n=1,\dots,4$ and $a=1,2,3$. From the definition \eqref{j} it follows that only
\begin{equation}
    J^{*a} = \r^{*\a_1\a_2} f_{\a_1 \a_2}{}^{\a_3}k_*{}^* k_{\a_3}{}^a = I^a ,
\end{equation}
must be kept. All the others vanish that is guaranteed by restricting to only bi-vector deformations, i.e. $r^{\a_1\a_2}=\rho^{*\a_1\a_2}$, $\r^{\a_1\a_2\a_3}=0$, and by setting  $f_{\a_1\a_2}{}^*=0$. Hence, 10D non-unimodular Yang-Baxter deformed backgrounds (of the appropriate 10=7+3 form) also solve our equations.

A natural question that arises here is whether the equations \eqref{eq:gengensugra0} are enough for the GS supermembrane to preserve $\kappa$-symmetry in analogy to 10D generalized supergravity equations, which ensure $\kappa$-symmetry of the GS superstring. As it has been discussed in the Introduction \ref{sec:intro} for the superstring the generalization is possible since a spinor superfield corresponding to the dimensions $\fr12$ of the supertorsion is no longer simply a spinorial derivative of the dilaton and contains an additional vector $I^\mu$. From the DFT point of view the vector $I^m$, that enters in $X^m$, is related to derivative of the dilaton w.r.t. a dual coordinate. A similar interpretation for the supermembrane lacks both the vector superfield and the dilaton to allow dependence on dual coordinates. The approach considered here resulting in \textbf{a} generalization of 11D supergravity, gives a clue to how to overcome the above obstacles. The main observation here is that ExFT is a theory with explicitly broken GL(11) symmetry as part of it enters the U-duality group. The diagonal GL(1) manifests itself in a scalar field $\phi$, which arises e.g. in the non-linear realization of the SL(5)$\times \RR^+$ U-duality symmetry \cite{Blair:2014zba}. A natural expectation would be that similar symmetry breaking in 11D would generate a proper superfield, that can be generalized to contain $J^{mn}$. A hint for why this could be possible is precisely the GL(11) symmetry breaking, which is the analogue of the Weyl-to-scale symmetry breaking for the RNS superstring on a generalized supergravity background. Most probably, to see the complete picture one should repeat our procedure for the full ExFT.

It is not yet clear whether there exists an interpretation of $J^{mn}$ in terms of derivatives along dual coordinates. The most naive expectation based on similarity of the truncated ExFT and DFT is that dual derivative of the field $\phi$ entering the generalized metric would give at least the antisymmetric part $J^{[mn]}\propto \tdt^{mn}\phi$. However, our preliminary analysis suggests
that this may not be enough as the external vielbein might also need to depend on dual coordinates to reproduce the corresponding shift of generalized flux. We plan to return to this issue soon.

Another interesting direction for further investigation is to ask what happens if one reduces the generalized supergravity equations obtained here to 10D as above but keeping tri-vector deformations, i.e. $\r^{\a_1\a_2\a_3}\neq 0$. This would give an extension of 10D generalized supergravity by adding tensor components $J^{ab}$. Naturally one is interested to see whether such equations ensure $\kappa$-symmetry of the superstring and whether such deformations preserve integrability of the 2D $\s$-model.

\section*{Acknowledgements}
This work has been supported by Russian Science Foundation grant RSCF-20-72-10144. 
NSD would like to acknowledge support from the ICTP through the Associates Programme (2017-2022) during the last phase of this paper.
\appendix
\setcounter{equation}{0}

\section{Ricci tensor and Ricci scalar in terms of anholonomicity coefficients}\label{appRicciGTR}

As we showed in this paper supergravity equations and their deformations can be derived from generalized flux formulation of extended field theories. To identify Einstein (and dilaton in the 10D case) equations in the standard form using this method we need to rewrite Ricci tensor and Ricci scalar in terms of anholonomicity coefficients (here referred to as fluxes) defined as usual as $[e_a,e_b] = f_{ab}{}^c e_c$. Since we keep these general and non-constant, $f_{ab}{}^c$ are convenient variables that encode geometric properties of the background. For the vielbein components $e_{a}{}^{m}$, $e^{a}{}_{m}$ we have
\begin{equation}
\begin{aligned}
f_{ab}{}^{c} = -2 e_{a}{}^{m} e_{b}{}^{n} \partial_{[m}e_{n]}{}^{c}, \qquad f_{a} = f_{ab}{}^{b} .
\end{aligned}
\end{equation}
All covariant objects built from vielbeins, such as Riemann and Ricci tensors and Ricci scalar, can be expressed in terms of these fluxes. To show that, we start with the expression for Christoffel symbols
\begin{equation}\label{eq:crist}
\begin{aligned}
\Gamma_{m n}{}^{k} = & \, \frac{1}{2}\, {h}^{k l} ({\partial}_{m}{{h}_{l n}}\,  + {\partial}_{n}{{h}_{l m}}\,  - {\partial}_{l}{{h}_{m n}}\, )  \\
= &  - \frac{1}{2}\, {e}^{a}\,_{m} {e}^{b}\,_{n} {e}^{c}\,_{l} {f}_{a c}\,^{d} {h}_{b d} {h}^{k l} - \frac{1}{2}\, {f}_{a b}\,^{c} {e}_{c}\,^{k} {e}^{a}\,_{m} {e}^{b}\,_{n} - \frac{1}{2}\, {e}^{a}\,_{m} {e}^{b}\,_{n} {e}^{c}\,_{l} {f}_{b c}\,^{d} {h}_{a d} {h}^{k l},
\end{aligned}
\end{equation}
and we obtain the Ricci tensor as
\begin{equation}\label{eq:riccitens}
\begin{aligned}
R_{n l} = & \,  {\partial}_{k}{{\Gamma}_{n l}\,^{k}}\,  - {\partial}_{n}{{\Gamma}_{k l}\,^{k}}\,  + {\Gamma}_{n l}\,^{p} {\Gamma}_{k p}\,^{k} - {\Gamma}_{k l}\,^{p} {\Gamma}_{n p}\,^{k}  \\
= &   - \frac{1}{2}\, {e}^{a}\,_{l} {e}^{b}\,_{n} {f}_{b c}\,^{d} {f}_{a}\,^{c}\,_{d} - \frac{1}{2}\, {\partial}_{m}{{f}_{a}\,^{b}\,_{c}}\,  {e}_{b}\,^{m} {e}^{c}\,_{l} {e}^{a}\,_{n} + \frac{1}{2}\, {e}_{a l} {e}^{b}\,_{n} {f}^{c} {f}_{b c}\,^{a} - \frac{1}{2}\, {e}^{a}\,_{l} {e}^{b}\,_{n} {f}_{a c}\,^{d} {f}_{b d}\,^{c} \\
& - \frac{1}{2}\, {\partial}_{m}{{f}_{a b}\,^{c}}\,  {e}_{c}\,^{m} {e}^{a}\,_{l} {e}^{b}\,_{n} - \frac{1}{2}\, {\partial}_{m}{{f}_{a}\,^{b}\,_{c}}\,  {e}_{b}\,^{m} {e}^{a}\,_{l} {e}^{c}\,_{n} + \frac{1}{2}\, {e}_{a n} {e}^{b}\,_{l} {f}^{c} {f}_{b c}\,^{a} + {\partial}_{n}{{f}_{a}}\,  {e}^{a}\,_{l} \\
& + \frac{1}{2}\, {e}^{a}\,_{l} {e}^{b}\,_{n} {f}_{c} {f}_{a b}\,^{c} + \frac{1}{4}\, {e}_{a l} {e}_{b n} {f}_{c d}\,^{b} {f}^{c d a},
\end{aligned}
\end{equation}
and the Ricci scalar as
\begin{equation}\label{eq:ricciscalar}
\begin{aligned}
R = & \,  h_{n l} ( {\partial}_{k}{{\Gamma}_{n l}\,^{k}}\,  - {\partial}_{n}{{\Gamma}_{k l}\,^{k}}\,  + {\Gamma}_{n l}\,^{p} {\Gamma}_{k p}\,^{k} - {\Gamma}_{k l}\,^{p} {\Gamma}_{n p}\,^{k} )  \\
= & - \frac{1}{2}\, {f}_{a b}\,^{c} {f}_{c d}\,^{a} {h}^{b d} - \frac{1}{4}\, {f}_{a b}\,^{c} {f}_{d f}\,^{g} {h}_{c g} {h}^{a d} {h}^{b f} + {\partial}_{m}{{f}_{a}}\,  {e}_{b}\,^{m} {h}^{a b} - {f}_{a} {f}_{b} {h}^{a b} + {\partial}_{m}{{f}_{a}}\,  {e}^{a}\,_{n} {h}^{m n}.
\end{aligned}
\end{equation}

\section{Linear part of the Bianchi identities on
\texorpdfstring{$J^{mn}$}{J}}
\label{app:BI_der}

Here we give details of the computation of the linear constraints on
$J^{mn}$ listed in \eqref{eq:condJ0} where Cadabra computer algebra with the code \texttt{BI\_to\_X} of \cite{gengensugra_2022} was used. To analyze constraints on $J^{mn}$ coming from Bianchi identities $Z_{MN,KL,P}{}^Q=0$ written in ``curved'' indices, it is convenient to decompose all expressions under the $\mathfrak{gl}(4)$ subalgebra of $\mathfrak{sl}(5)$.  For that we first split the index $M=(5,m)$ and list all non-vanishing components of $Z_{MN,KL,P}{}^Q$:
\begin{equation}
    \label{eq:allJ}
    \begin{aligned}
        Z_{[12,34,5]}{}^l=&\   {\partial}_{p}{{J}^{p l}}\, +\, {J}^{p q}  {\partial}_{p}{{e}_{q}\,^{a}}\,  {e}_{a}\,^{l} - \, {J}^{p l}  {\partial}_{p}{\phi}\,  + {J}^{p l}  {\partial}_{p}{{e}_{q}\,^{a}}\,  {e}_{a}\,^{q};\\
        Z_{[mn],[kl]}{}^p=&\ {\delta}_{m}\,^{p} ( - {\epsilon}_{n k l q} {\partial}_{r}{{J}^{r q}}\,  + 2\, {J}^{q r} {\epsilon}_{n r k l} {\partial}_{q}{\phi}\,  - {J}^{q r} {\epsilon}_{n k l s} {\partial}_{q}{{e}_{r}\,^{a}}\,  {e}_{a}\,^{s} - {J}^{q r} {\epsilon}_{n r k l} {\partial}_{q}{{e}_{s}\,^{a}}\,  {e}_{a}\,^{s}) \\
        &- {\epsilon}_{m k l q} {\partial}_{n}{{J}^{p q}}\,  - {J}^{p q} {\epsilon}_{m k l r} {\partial}_{n}{{e}_{q}\,^{a}}\,  {e}_{a}\,^{r} + {J}^{p q} {\epsilon}_{m q k l} {\partial}_{n}{\phi}\,  + {J}^{p q} {\epsilon}_{m n q r} {\partial}_{k}{{e}_{l}\,^{a}}\,  {e}_{a}\,^{r} + {J}^{p q} {\epsilon}_{m n q k} {\partial}_{l}{{e}_{r}\,^{a}}\,  {e}_{a}\,^{r} \\
        &- {J}^{p q} {\epsilon}_{m n q k} {\partial}_{r}{{e}_{l}\,^{a}}\,  {e}_{a}\,^{r} - {J}^{p q} {\epsilon}_{m n q k} {\partial}_{l}{\phi}\,  - {J}^{p q} {\epsilon}_{q k l r} {\partial}_{m}{{e}_{n}\,^{a}}\,  {e}_{a}\,^{r} - {J}^{q r} {\epsilon}_{m r k l} {\partial}_{q}{{e}_{n}\,^{a}}\,  {e}_{a}\,^{p};\\
        Z_{m,l,[nk]}{}^p=&\ {\delta}_{l}\,^{p} (\frac{1}{2}\, {\epsilon}_{m n k q} {\partial}_{r}{{J}^{r q}}\,  - \frac{3}{2}\, {J}^{q r} {\epsilon}_{m r n k} {\partial}_{q}{\phi}\,  + \frac{1}{2}\, {J}^{q r} {\epsilon}_{m n k s} {\partial}_{q}{{e}_{r}\,^{a}}\,  {e}_{a}\,^{s} + \frac{1}{2}\, {J}^{q r} {\epsilon}_{m r n k} {\partial}_{q}{{e}_{s}\,^{a}}\,  {e}_{a}\,^{s}) \\
        &- {\delta}_{m}\,^{p} {J}^{q r} {\epsilon}_{r l n k} {\partial}_{q}{\phi}\,
        + {\epsilon}_{l n k q} {\partial}_{m}{{J}^{p q}}\,  - {\epsilon}_{m n k q} {\partial}_{l}{{J}^{p q}}\,  + {J}^{p q} {\epsilon}_{l n k r} {\partial}_{m}{{e}_{q}\,^{a}}\,  {e}_{a}\,^{r} + {J}^{p q} {\epsilon}_{q l n k} {\partial}_{m}{\phi}\, \\
        &- {J}^{p q} {\epsilon}_{m n k r} {\partial}_{l}{{e}_{q}\,^{a}}\,  {e}_{a}\,^{r} + {J}^{p q} {\epsilon}_{m q n k} {\partial}_{l}{\phi}\, 
        - 2\, {J}^{p q} {\epsilon}_{m q l r} {\partial}_{n}{{e}_{k}\,^{a}}\,  {e}_{a}\,^{r} - 2\, {J}^{p q} {\epsilon}_{m q l n} {\partial}_{k}{{e}_{r}\,^{a}}\,  {e}_{a}\,^{r} \\
        &+ 2\, {J}^{p q} {\epsilon}_{m q l n} {\partial}_{r}{{e}_{k}\,^{a}}\,  {e}_{a}\,^{r} + 2\, {J}^{p q} {\epsilon}_{m q l n} {\partial}_{k}{\phi}\,  - {J}^{p q} {\epsilon}_{q n k r} {\partial}_{m}{{e}_{l}\,^{a}}\,  {e}_{a}\,^{r} 
        + {J}^{p q} {\epsilon}_{q n k r} {\partial}_{l}{{e}_{m}\,^{a}}\,  {e}_{a}\,^{r}\\
        &- {J}^{q r} {\epsilon}_{m r n k} {\partial}_{q}{{e}_{l}\,^{a}}\,  {e}_{a}\,^{p} - {J}^{q r} {\epsilon}_{r l n k} {\partial}_{q}{{e}_{m}\,^{a}}\,  {e}_{a}\,^{p} ;\\
        Z_{[mn],k,l}{}^p=&\ {\delta}_{k}\,^{p} {J}^{q r} {\epsilon}_{m n r l} {\partial}_{q}{\phi}\,  - {\epsilon}_{m n l q} {\partial}_{k}{{J}^{p q}}\,  - 2\, {J}^{p q} {\epsilon}_{m q l r} {\partial}_{n}{{e}_{k}\,^{a}}\,  {e}_{a}\,^{r} + 2\, {J}^{p q} {\epsilon}_{m q l r} {\partial}_{k}{{e}_{n}\,^{a}}\,  {e}_{a}\,^{r} \\
        &- {J}^{p q} {\epsilon}_{m n q l} {\partial}_{k}{{e}_{r}\,^{a}}\,  {e}_{a}\,^{r}+ {J}^{p q} {\epsilon}_{m n q l} {\partial}_{r}{{e}_{k}\,^{a}}\,  {e}_{a}\,^{r} + {J}^{p q} {\epsilon}_{m n q l} {\partial}_{k}{\phi}\,  - 2\, {J}^{p q} {\epsilon}_{m q l k} {\partial}_{n}{\phi}\, \\
        &- {J}^{p q} {\epsilon}_{m n q r} {\partial}_{l}{{e}_{k}\,^{a}}\,  {e}_{a}\,^{r} + {J}^{p q} {\epsilon}_{m n q r} {\partial}_{k}{{e}_{l}\,^{a}}\,  {e}_{a}\,^{r} - {J}^{p q} {\epsilon}_{m n q k} {\partial}_{l}{\phi}\,  - {J}^{p q} {\epsilon}_{m n l r} {\partial}_{k}{{e}_{q}\,^{a}}\,  {e}_{a}\,^{r} \\
        &+ {J}^{q r} {\epsilon}_{m n r l} {\partial}_{q}{{e}_{k}\,^{a}}\,  {e}_{a}\,^{p}.
    \end{aligned}
\end{equation}
These in general belong to reducible representations of $\mathfrak{gl}(4)$ and can be further decomposed. The first line $Z^m \in \bf 4$ is already an irrep, hence, we proceed with the second line:
\begin{equation}
    Z_{[mn],[kl]}{}^p \in {\bf 6\times 6 \times \bar{4} } \to ({\bf 1+15+20' })\times \bf \bar{4}.
\end{equation}
Direct check shows that for the given $X_{ABC}{}^D$ the part in $\bf 20'$ identically vanishes, leaving us with
\begin{equation}
    Z_{[mn],[kl]}{}^p \in ({\bf 1+15})\times {\bf \bar{4} } = \bf \bar{4} + \bar{4} + \ol{20} + \ol{36}.
\end{equation}
In tensor components these can be written as follows
\begin{equation}
    \begin{aligned}
    & \bf \bar{4}: && \e^{mnkl}Z_{mnkl}{}^p,\\
    & \bf \bar{4}: && \e^{mnkl}Z_{mnkp}{}^p - \fr14 \e^{mnkp}Z_{mnkp}{}^l,\\
    & \bf \ol{20}: && \e^{mnk[p}Z_{mnkl}{}^{q]} - tr,\\
    & \bf \ol{36}: && \e^{mnk(p}Z_{mnkl}{}^{q)} - tr.
    \end{aligned}
\end{equation}
where the trace parts include the irreps in the first two lines and hence vanish. Explicitly these give the following constraints
\begin{equation}
    \begin{aligned}
    & \bf \bar{4}: &&     \dt_n I^{nm} - 2 I^{nm}\dt_n \phi + I^{nm}e^{-1}\dt_n e=0,\\
    & \bf \bar{4}: &&    - 4\, {U}_{m} {\partial}_{n}{{J}^{n m}}\,  + 12\, {U}_{m} {J}^{n m} {d}_{n} - 8\, {U}_{m} {J}^{n k} {\partial}_{n}{{e}_{k}\,^{a}}\,  {e}_{a}\,^{m} - 4\, {U}_{m} {J}^{n m} {\partial}_{n}{{e}_{k}\,^{a}}\,  {e}_{a}\,^{k} - 2\, {U}_{m} {\partial}_{n}{{J}^{m n}}\, \\
    & &&+ 3\, {U}_{m} {J}^{m n} {d}_{n} + 2\, {U}_{m} {J}^{n k} {\partial}_{k}{{e}_{n}\,^{a}}\,  {e}_{a}\,^{m} - 2\, {U}_{m} {J}^{m n} {\partial}_{n}{{e}_{k}\,^{a}}\,  {e}_{a}\,^{k}; \\
    & \bf \ol{20}: &&  {A}_{m n} {U}^{k} ( - {J}^{m l} {\partial}_{l}{{e}_{k}\,^{a}}\,  {e}_{a}\,^{n} + {J}^{l m} {\partial}_{l}{{e}_{k}\,^{a}}\,  {e}_{a}\,^{n} - {\partial}_{k}{{J}^{m n}}\, )\\
    & &&+ {A}_{m n} {U}^{m} ({J}^{k l} {\partial}_{k}{{e}_{l}\,^{a}}\,  {e}_{a}\,^{n} - {J}^{k l} {\partial}_{l}{{e}_{k}\,^{a}}\,  {e}_{a}\,^{n}) + 6\, {A}_{m n} {U}^{n} {J}^{k m} {d}_{k}=0,\\
    & \bf \ol{36}: && {S}_{m n} {U}^{k} (2\, {J}^{m l} {\partial}_{l}{{e}_{k}\,^{a}}\,  {e}_{a}\,^{n} + 2\, {\partial}_{k}{{J}^{m n}}\,  + 2\, {J}^{l m} {\partial}_{l}{{e}_{k}\,^{a}}\,  {e}_{a}\,^{n}) + {S}_{m n} {U}^{n} ( - {J}^{m k} {d}_{k} + 2\, {J}^{k m} {d}_{k});=0,
    \end{aligned}
\end{equation}
where $J^{mn} = I^{mn} + S^{mn}$. Note that expressions in the third and the fourth line are not irreducible representations, but some combinations of that with the trace part.  

Consider now $Z_{[mn],k,l}{}^p$, whose lower index structure decomposes as
\begin{equation}
        Z_{[mn],k,l} \in {\bf 6\times 4 \times 4 } \to  {\bf 6\times ( 6 + {10})} = ({\bf 1+15+20' }) + \bf (15 + {45}).
\end{equation}
As before, direct check shows that the $\bf 20'$ and $\bf {45}$ identically vanish. Hence, in total we have $Z_{[mn],k,l}{}^p \in ({\bf 1 + 15 + 15  } )\times \bf \bar{4}$. Before proceeding with decomposing these irrep products we notice that the difference between the two $\bf 15 \times \bar{4}$'s can be represented in the following nice form
\begin{equation}
    \dt_m J^{k l} + J^{k n}\dt_{n}e_m{}^a e_a{}^l+ J^{n l}\dt_{n}e_m{}^a e_a{}^k + J^{k n}\d_m{}^l \dt_n \phi+ J^{n l}\d_m{}^k \dt_n \phi =0,
\end{equation}
given which it is enough to keep only one of these representations. Next we consider the $\bf \bar{4}$'s, which given the previous conditions both boil down to
\begin{equation}
    J^{mn}\dt_n \phi=0.
\end{equation}
Given these conditions the remaining parts in the $\bf \ol{20} + \ol{36}$ are equivalent to the previous constraints. The same is true for  $Z_{m,n,[kl]}{}^p$. 
Collecting all the conditions together we have
\begin{equation}
    \begin{aligned}
        &1.&&J^{mn}\dt_n \phi=0,\\
        &2.&&{\partial}_{p}{{J}^{p l}}\, +\, {J}^{p q}  {\partial}_{p}{{e}_{q}\,^{a}}\,  {e}_{a}\,^{l} - \, {J}^{p l}  {\partial}_{p}{\phi}\,  + {J}^{p l}  {\partial}_{p}{{e}_{q}\,^{a}}\,  {e}_{a}\,^{q}=0 ,\\
        &3.&&     \dt_n I^{nm} - 2 I^{nm}\dt_n \phi + I^{nm}e^{-1}\dt_n e=0,\\
        &4.&&    - 2\, {U}_{m} {\partial}_{n}{{J}^{n m}}\,  + 6\, {U}_{m} {J}^{n m} {d}_{n} - 4\, {U}_{m} {J}^{n k} {\partial}_{n}{{e}_{k}\,^{a}}\,  {e}_{a}\,^{m} - 2\, {U}_{m} {J}^{n m} {\partial}_{n}{{e}_{k}\,^{a}}\,  {e}_{a}\,^{k} - \, {U}_{m} {\partial}_{n}{{J}^{m n}}\, \\
        & &&+\, {U}_{m} {J}^{n k} {\partial}_{k}{{e}_{n}\,^{a}}\,  {e}_{a}\,^{m} - \, {U}_{m} {J}^{m n} {\partial}_{n}{{e}_{k}\,^{a}}\,  {e}_{a}\,^{k}=0 , \\ 
       &5. &&\dt_m J^{k l} + J^{k n}\dt_{n}e_m{}^a e_a{}^l+ J^{n l}\dt_{n}e_m{}^a e_a{}^k + J^{n l}\d_m{}^k \dt_n \phi =0,\\
         &6.&&  {A}_{m n} {U}^{k} ( - {J}^{m l} {\partial}_{l}{{e}_{k}\,^{a}}\,  {e}_{a}\,^{n} + {J}^{l m} {\partial}_{l}{{e}_{k}\,^{a}}\,  {e}_{a}\,^{n} - {\partial}_{k}{{J}^{m n}}\, )\\
       &&&+ {A}_{m n} {U}^{m} ({J}^{k l} {\partial}_{k}{{e}_{l}\,^{a}}\,  {e}_{a}\,^{n} - {J}^{k l} {\partial}_{l}{{e}_{k}\,^{a}}\,  {e}_{a}\,^{n}) + 6\, {A}_{m n} {U}^{n} {J}^{k m} {d}_{k}=0,\\
        &7.&& {S}_{m n} {U}^{k} (\, {J}^{m l} {\partial}_{l}{{e}_{k}\,^{a}}\,  {e}_{a}\,^{n} +  {\partial}_{k}{{J}^{m n}}\,  +  {J}^{l m} {\partial}_{l}{{e}_{k}\,^{a}}\,  {e}_{a}\,^{n}) + {S}_{m n} {U}^{n} {J}^{k m} {d}_{k}=0 .
    \end{aligned}
\end{equation}
A combination of 2, 3 and 4 gives
\begin{equation}
     8.\quad - {U}_{k} {J}^{m n} {\partial}_{m}{{e}_{n}\,^{a}}\,  {e}_{a}\,^{k} + 5\, {U}_{m} {J}^{n m} {d}_{n} + {U}_{k} {J}^{m n} {\partial}_{n}{{e}_{m}\,^{a}}\,  {e}_{a}\,^{k}=0 .
\end{equation}
Sum of 6 and 7 together with 8 gives 5. Trace of 5 together with 8 gives 2. Trace of 6 together with 8 gives 3. Hence, the only independent conditions are
\begin{equation}
    \label{eq:condJ1}
    \begin{aligned}
        0&=\dt_m J^{k l} + J^{k n}\dt_{n}e_m{}^a e_a{}^l+ J^{n l}\dt_{n}e_m{}^a e_a{}^k + J^{n l}\d_m{}^k \dt_n \phi,\\
        0&=I^{mn}\dt_m e_n{}^a e_a{}^k - \fr52 J^{lk}\dt_l \phi \quad \longrightarrow    \nabla_m\big(e^{-\f}I^{mn}\big) = 0 ,\\
        0&=J^{mn}\dt_n \phi.
    \end{aligned}
\end{equation}

Terms in Bianchi identities that are proportional to the gauge field $V^m$ appear independently from the above. Following the same steps we arrive at
\begin{equation}
    \begin{aligned}
        \nabla_{[m}Z_{n]}  - \fr13 J^{kl}F_{mnkl}&=0,\\
        \nabla_k\Big(e^{-\phi}J^{k[l}V^{p]}\Big) &=0,\\
        \nabla_k(J^{(pl)}V^k) - \nabla_k(V^{(p}J^{l)k})&=0.
    \end{aligned}
\end{equation}
Interestingly enough, the first condition ensures that equations following from the  antisymmetric part of $\delta e_m{}^a$ are satisfied identically.

\bibliography{bib.bib}

\providecommand{\href}[2]{#2}\begingroup\raggedright\begin{thebibliography}{10}

\bibitem{Hull:1994ys}
C.~Hull and P.~Townsend, ``{Unity of superstring dualities},''
  \href{http://dx.doi.org/10.1016/0550-3213(94)00559-W}{{\em Nucl.Phys.}
  {\bfseries B438} (1995) 109--137},
  \href{http://arxiv.org/abs/hep-th/9410167}{{\ttfamily arXiv:hep-th/9410167
  [hep-th]}}.

\bibitem{Witten:1995ex}
E.~Witten, ``{String theory dynamics in various dimensions},''
  \href{http://dx.doi.org/10.1016/0550-3213(95)00158-O}{{\em Nucl.Phys.}
  {\bfseries B443} (1995) 85--126},
  \href{http://arxiv.org/abs/hep-th/9503124}{{\ttfamily arXiv:hep-th/9503124
  [hep-th]}}.

\bibitem{Townsend:1996xj}
P.~K. Townsend, ``{Four lectures on M theory},'' in {\em {ICTP Summer School in
  High-energy Physics and Cosmology}}, pp.~385--438.
\newblock 12, 1996.
\newblock \href{http://arxiv.org/abs/hep-th/9612121}{{\ttfamily
  arXiv:hep-th/9612121}}.

\bibitem{Duff:1996zn}
M.~J. Duff, ``{Supermembranes},'' in {\em {Theoretical Advanced Study Institute
  in Elementary Particle Physics (TASI 96): Fields, Strings, and Duality}}.
\newblock 6, 1996.
\newblock \href{http://arxiv.org/abs/hep-th/9611203}{{\ttfamily
  arXiv:hep-th/9611203}}.

\bibitem{Howe:1977hp}
P.~S. Howe and R.~W. Tucker, ``{A Locally Supersymmetric and Reparametrization
  Invariant Action for a Spinning Membrane},''
  \href{http://dx.doi.org/10.1088/0305-4470/10/9/003}{{\em J. Phys. A}
  {\bfseries 10} (1977) L155--L158}.

\bibitem{PhysRevD.3.2415}
P.~Ramond, ``Dual Theory for Free Fermions,''
  \href{http://dx.doi.org/10.1103/PhysRevD.3.2415}{{\em Phys. Rev. D}
  {\bfseries 3} (May, 1971) 2415--2418}.

\bibitem{NEVEU197186}
A.~Neveu and J.~Schwarz, ``Factorizable dual model of pions,''
  \href{http://dx.doi.org/https://doi.org/10.1016/0550-3213(71)90448-2}{{\em
  Nucl. Phys. B} {\bfseries 31} no.~1, (1971) 86--112}.

\bibitem{FRIEDAN198693}
D.~Friedan, E.~Martinec, and S.~Shenker, ``Conformal invariance, supersymmetry
  and string theory,''
  \href{http://dx.doi.org/10.1016/S0550-3213(86)80006-2}{{\em Nuclear Physics
  B} {\bfseries 271} no.~3, (1986) 93--165}.

\bibitem{Deser1976ACA}
S.~Deser and B.~Zumino, ``A complete action for the spinning string,''
  \href{http://dx.doi.org/10.1016/0370-2693(76)90245-8}{{\em Phys. Lett. B}
  {\bfseries 65} (1976) 369--373}.

\bibitem{Brink:1976sc}
L.~Brink, P.~Di~Vecchia, and P.~S. Howe, ``{A Locally Supersymmetric and
  Reparametrization Invariant Action for the Spinning String},''
  \href{http://dx.doi.org/10.1016/0370-2693(76)90445-7}{{\em Phys. Lett. B}
  {\bfseries 65} (1976) 471--474}.

\bibitem{Bergshoeff:1987cm}
E.~Bergshoeff, E.~Sezgin, and P.~K. Townsend, ``{Supermembranes and
  Eleven-Dimensional Supergravity},''
\href{http://dx.doi.org/10.1016/0370-2693(87)91272-X}{{\em Phys. Lett.}
  {\bfseries B189} (1987) 75--78}.

\bibitem{Bergshoeff:1987qx}
E.~Bergshoeff, E.~Sezgin, and P.~K. Townsend, ``{Properties of the
  Eleven-Dimensional Super Membrane Theory},''
  \href{http://dx.doi.org/10.1016/0003-4916(88)90050-4}{{\em Annals Phys.}
  {\bfseries 185} (1988) 330}.

\bibitem{Green:1983wt}
M.~B. Green and J.~H. Schwarz, ``{Covariant Description of Superstrings},''
  \href{http://dx.doi.org/10.1016/0370-2693(84)92021-5}{{\em Phys. Lett. B}
  {\bfseries 136} (1984) 367--370}.

\bibitem{Green:1987sp}
M.~B. Green, J.~Schwarz, and E.~Witten, {\em {Superstring theory. Vol. 1:
  Introduction}}.
\newblock Cambridge University Press, Cambridge, England, 1987.

\bibitem{Bandos:1995zw}
I.~A. Bandos, D.~P. Sorokin, M.~Tonin, P.~Pasti, and D.~V. Volkov,
  ``{Superstrings and supermembranes in the doubly supersymmetric geometrical
  approach},'' \href{http://dx.doi.org/10.1016/0550-3213(95)00267-V}{{\em Nucl.
  Phys. B} {\bfseries 446} (1995) 79--118},
  \href{http://arxiv.org/abs/hep-th/9501113}{{\ttfamily arXiv:hep-th/9501113}}.

\bibitem{Sorokin:1999jx}
D.~P. Sorokin, ``{Superbranes and superembeddings},''
  \href{http://dx.doi.org/10.1016/S0370-1573(99)00104-0}{{\em Phys. Rept.}
  {\bfseries 329} (2000) 1--101},
  \href{http://arxiv.org/abs/hep-th/9906142}{{\ttfamily arXiv:hep-th/9906142}}.

\bibitem{Callan:1989nz}
C.~G. Callan, Jr. and L.~Thorlacius, ``{Sigma Models and String Theory},'' in
  {\em {Theoretical Advanced Study Institute in Elementary Particle Physics:
  Particles, Strings and Supernovae (TASI 88)}}.
\newblock 3, 1989.

\bibitem{CALLAN1987525}
C.~Callan, C.~Lovelace, C.~Nappi, and S.~Yost, ``String loop corrections to
  beta functions,''
  \href{http://dx.doi.org/https://doi.org/10.1016/0550-3213(87)90227-6}{{\em
  Nucl. Phys. B} {\bfseries 288} (1987) 525--550}.

\bibitem{CALLAN1985593}
C.~Callan, D.~Friedan, E.~Martinec, and M.~Perry, ``Strings in background
  fields,''
  \href{http://dx.doi.org/https://doi.org/10.1016/0550-3213(85)90506-1}{{\em
  Nucl. Phys. B} {\bfseries 262} no.~4, (1985) 593--609}.

\bibitem{Callan:1986bc}
C.~G. Callan, Jr., C.~Lovelace, C.~R. Nappi, and S.~A. Yost, ``{String Loop
  Corrections to beta Functions},''
  \href{http://dx.doi.org/10.1016/0550-3213(87)90227-6}{{\em Nucl. Phys. B}
  {\bfseries 288} (1987) 525--550}.

\bibitem{Callan:1985ia}
C.~G. {Callan, Jr.}, E.~Martinec, M.~Perry, and D.~Friedan, ``{Strings in
  background fields},''
  \href{http://dx.doi.org/10.1016/0550-3213(85)90506-1}{{\em Nucl.Phys.}
  {\bfseries B262} (1985) 593}.

\bibitem{Wulff:2016tju}
L.~Wulff and A.~A. Tseytlin, ``{Kappa-symmetry of superstring sigma model and
  generalized 10d supergravity equations},''
  \href{http://dx.doi.org/10.1007/JHEP06(2016)174}{{\em JHEP} {\bfseries 06}
  (2016) 174},
\href{http://arxiv.org/abs/1605.04884}{{\ttfamily arXiv:1605.04884 [hep-th]}}.

\bibitem{Arutyunov:2015mqj}
G.~Arutyunov, S.~Frolov, B.~Hoare, R.~Roiban, and A.~A. Tseytlin, ``{Scale
  invariance of the $\eta$-deformed $AdS_5\times S^5$ superstring, T-duality
  and modified type II equations},''
  \href{http://dx.doi.org/10.1016/j.nuclphysb.2015.12.012}{{\em Nucl. Phys.}
  {\bfseries B903} (2016) 262--303},
\href{http://arxiv.org/abs/1511.05795}{{\ttfamily arXiv:1511.05795 [hep-th]}}.

\bibitem{Fernandez-Melgarejo:2018wpg}
J.~J. Fernández-Melgarejo, J.-I. Sakamoto, Y.~Sakatani, and K.~Yoshida,
  ``{Weyl invariance of string theories in generalized supergravity
  backgrounds},'' \href{http://dx.doi.org/10.1103/PhysRevLett.122.111602}{{\em
  Phys. Rev. Lett.} {\bfseries 122} no.~11, (2019) 111602},
\href{http://arxiv.org/abs/1811.10600}{{\ttfamily arXiv:1811.10600 [hep-th]}}.

\bibitem{Muck:2019pwj}
W.~M\"uck, ``{Generalized Supergravity Equations and Generalized
  Fradkin-Tseytlin Counterterm},''
  \href{http://dx.doi.org/10.1007/JHEP05(2019)063}{{\em JHEP} {\bfseries 05}
  (2019) 063}, \href{http://arxiv.org/abs/1904.06126}{{\ttfamily
  arXiv:1904.06126 [hep-th]}}.

\bibitem{Cherednik:1981df}
I.~V. Cherednik, ``{Relativistically Invariant Quasiclassical Limits of
  Integrable Two-dimensional Quantum Models},''
  \href{http://dx.doi.org/10.1007/BF01086395}{{\em Theor. Math. Phys.}
  {\bfseries 47} (1981) 422--425}.

\bibitem{Klimcik:2002zj}
C.~Klim\v{c}{\'i}k, ``{Yang-Baxter sigma models and dS/AdS T duality},''
  \href{http://dx.doi.org/10.1088/1126-6708/2002/12/051}{{\em JHEP} {\bfseries
  12} (2002) 051},
\href{http://arxiv.org/abs/hep-th/0210095}{{\ttfamily arXiv:hep-th/0210095
  [hep-th]}}.

\bibitem{Klimcik:2008eq}
C.~Klim\v{c}{\'i}k, ``{On integrability of the Yang-Baxter sigma-model},''
  \href{http://dx.doi.org/10.1063/1.3116242}{{\em J. Math. Phys.} {\bfseries
  50} (2009) 043508},
\href{http://arxiv.org/abs/0802.3518}{{\ttfamily arXiv:0802.3518 [hep-th]}}.

\bibitem{Delduc:2013fga}
F.~Delduc, M.~Magro, and B.~Vicedo, ``{On classical $q$-deformations of
  integrable sigma-models},''
  \href{http://dx.doi.org/10.1007/JHEP11(2013)192}{{\em JHEP} {\bfseries 11}
  (2013) 192},
\href{http://arxiv.org/abs/1308.3581}{{\ttfamily arXiv:1308.3581 [hep-th]}}.

\bibitem{Bena:2003wd}
I.~Bena, J.~Polchinski, and R.~Roiban, ``{Hidden symmetries of the AdS(5) x
  S**5 superstring},'' \href{http://dx.doi.org/10.1103/PhysRevD.69.046002}{{\em
  Phys. Rev. D} {\bfseries 69} (2004) 046002},
  \href{http://arxiv.org/abs/hep-th/0305116}{{\ttfamily arXiv:hep-th/0305116}}.

\bibitem{Arutyunov:2015qva}
G.~Arutyunov, R.~Borsato, and S.~Frolov, ``{Puzzles of $\eta$-deformed AdS$_5
  \times$ S$^5$},'' \href{http://dx.doi.org/10.1007/JHEP12(2015)049}{{\em JHEP}
  {\bfseries 12} (2015) 049},
\href{http://arxiv.org/abs/1507.04239}{{\ttfamily arXiv:1507.04239 [hep-th]}}.

\bibitem{Hoare:2015wia}
B.~Hoare and A.~A. Tseytlin, ``{Type IIB supergravity solution for the T-dual
  of the $\eta$-deformed AdS$_{5} \times$ S$^{5}$ superstring},''
  \href{http://dx.doi.org/10.1007/JHEP10(2015)060}{{\em JHEP} {\bfseries 10}
  (2015) 060},
\href{http://arxiv.org/abs/1508.01150}{{\ttfamily arXiv:1508.01150 [hep-th]}}.

\bibitem{Hoare:2015gda}
B.~Hoare and A.~A. Tseytlin, ``{On integrable deformations of superstring sigma
  models related to $AdS_n \times S^n$ supercosets},''
  \href{http://dx.doi.org/10.1016/j.nuclphysb.2015.06.001}{{\em Nucl. Phys.}
  {\bfseries B897} (2015) 448--478},
\href{http://arxiv.org/abs/1504.07213}{{\ttfamily arXiv:1504.07213 [hep-th]}}.

\bibitem{Hughes:1986fa}
J.~Hughes, J.~Liu, and J.~Polchinski, ``{Supermembranes},''
  \href{http://dx.doi.org/10.1016/0370-2693(86)91204-9}{{\em Phys. Lett. B}
  {\bfseries 180} (1986) 370--374}.

\bibitem{Howe:1997he}
P.~S. Howe, ``{Weyl superspace},''
  \href{http://dx.doi.org/10.1016/S0370-2693(97)01261-6}{{\em Phys. Lett.}
  {\bfseries B415} (1997) 149--155},
\href{http://arxiv.org/abs/hep-th/9707184}{{\ttfamily arXiv:hep-th/9707184
  [hep-th]}}.

\bibitem{Bakhmatov:2022rjn}
I.~Bakhmatov, A.~\c{C}atal \"Ozer, N.~S. Deger, K.~Gubarev, and E.~T. Musaev,
  ``{Generalizing eleven-dimensional supergravity},''
  \href{http://dx.doi.org/10.1103/PhysRevD.105.L081904}{{\em Phys. Rev. D}
  {\bfseries 105} no.~8, (2022) L081904},
  \href{http://arxiv.org/abs/2203.03372}{{\ttfamily arXiv:2203.03372
  [hep-th]}}.

\bibitem{Bakhmatov:2017joy}
I.~Bakhmatov, O.~Kelekci, E.~\'O~Colg\'ain, and M.~M. Sheikh-Jabbari,
  ``{Classical Yang-Baxter Equation from Supergravity},''
  \href{http://dx.doi.org/10.1103/PhysRevD.98.021901}{{\em Phys. Rev.}
  {\bfseries D98} no.~2, (2018) 021901},
\href{http://arxiv.org/abs/1710.06784}{{\ttfamily arXiv:1710.06784 [hep-th]}}.

\bibitem{Bakhmatov:2018apn}
I.~Bakhmatov, E.~\'O~Colg\'ain, M.~M. Sheikh-Jabbari, and H.~Yavartanoo,
  ``{Yang-Baxter Deformations Beyond Coset Spaces (a slick way to do TsT)},''
  \href{http://dx.doi.org/10.1007/JHEP06(2018)161}{{\em JHEP} {\bfseries 06}
  (2018) 161},
\href{http://arxiv.org/abs/1803.07498}{{\ttfamily arXiv:1803.07498 [hep-th]}}.

\bibitem{Borsato:2020bqo}
R.~Borsato, A.~Vilar~L\'opez, and L.~Wulff, ``{The first $\alpha'$-correction
  to homogeneous Yang-Baxter deformations using $O(d, d)$},''
  \href{http://dx.doi.org/10.1007/JHEP07(2020)103}{{\em JHEP} {\bfseries 07}
  no.~07, (2020) 103}, \href{http://arxiv.org/abs/2003.05867}{{\ttfamily
  arXiv:2003.05867 [hep-th]}}.

\bibitem{Gubarev:2020ydf}
K.~Gubarev and E.~T. Musaev, ``{Polyvector deformations in eleven-dimensional
  supergravity},'' \href{http://dx.doi.org/10.1103/PhysRevD.103.066021}{{\em
  Phys. Rev. D} {\bfseries 103} no.~6, (2021) 066021},
  \href{http://arxiv.org/abs/2011.11424}{{\ttfamily arXiv:2011.11424
  [hep-th]}}.

\bibitem{Aldazabal:2013sca}
G.~Aldazabal, D.~Marqu{\'e}s, and C.~N{\'u}{\~n}ez, ``{Double Field Theory: A
  Pedagogical Review},''
  \href{http://dx.doi.org/10.1088/0264-9381/30/16/163001}{{\em Class. Quant.
  Grav.} {\bfseries 30} (2013) 163001},
\href{http://arxiv.org/abs/1305.1907}{{\ttfamily arXiv:1305.1907 [hep-th]}}.

\bibitem{Berman:2013eva}
D.~S. Berman and D.~C. Thompson, ``{Duality Symmetric String and M-Theory},''
  \href{http://dx.doi.org/10.1016/j.physrep.2014.11.007}{{\em Phys. Rept.}
  {\bfseries 566} (2014) 1--60},
\href{http://arxiv.org/abs/1306.2643}{{\ttfamily arXiv:1306.2643 [hep-th]}}.

\bibitem{Hohm:2013bwa}
O.~Hohm, D.~L{\"u}st, and B.~Zwiebach, ``{The Spacetime of Double Field Theory:
  Review, Remarks, and Outlook},''
  \href{http://dx.doi.org/10.1002/prop.201300024}{{\em Fortsch. Phys.}
  {\bfseries 61} (2013) 926--966},
\href{http://arxiv.org/abs/1309.2977}{{\ttfamily arXiv:1309.2977 [hep-th]}}.

\bibitem{Baguet:2015xha}
A.~Baguet, O.~Hohm, and H.~Samtleben, ``{E$_{6(6)}$ Exceptional Field Theory:
  Review and Embedding of Type IIB},'' {\em PoS} {\bfseries CORFU2014} (2015)
  133,
\href{http://arxiv.org/abs/1506.01065}{{\ttfamily arXiv:1506.01065 [hep-th]}}.

\bibitem{Musaev:2019zcr}
E.~T. Musaev, ``{U-Dualities in Type II and M-Theory: A Covariant Approach},''
  \href{http://dx.doi.org/10.3390/sym11080993}{{\em Symmetry} {\bfseries 11}
  no.~8, (2019) 993}.

\bibitem{Berman:2020tqn}
D.~S. Berman and C.~D. Blair, ``{The Geometry, Branes and Applications of
  Exceptional Field Theory},''
  \href{http://dx.doi.org/10.1142/S0217751X20300148}{{\em Int. J. Mod. Phys. A}
  {\bfseries 35} no.~30, (6, 2020) 2030014},
  \href{http://arxiv.org/abs/2006.09777}{{\ttfamily arXiv:2006.09777
  [hep-th]}}.

\bibitem{vanTongeren:2015uha}
S.~J. van Tongeren, ``{Yang–Baxter deformations, AdS/CFT, and
  twist-noncommutative gauge theory},''
  \href{http://dx.doi.org/10.1016/j.nuclphysb.2016.01.012}{{\em Nucl. Phys.}
  {\bfseries B904} (2016) 148--175},
\href{http://arxiv.org/abs/1506.01023}{{\ttfamily arXiv:1506.01023 [hep-th]}}.

\bibitem{Hoare:2016hwh}
B.~Hoare and S.~J. van Tongeren, ``{On jordanian deformations of AdS$_5$ and
  supergravity},'' \href{http://dx.doi.org/10.1088/1751-8113/49/43/434006}{{\em
  J. Phys. A} {\bfseries 49} no.~43, (2016) 434006},
  \href{http://arxiv.org/abs/1605.03554}{{\ttfamily arXiv:1605.03554
  [hep-th]}}.

\bibitem{Orlando:2016qqu}
D.~Orlando, S.~Reffert, J.-i. Sakamoto, and K.~Yoshida, ``{Generalized type IIB
  supergravity equations and non-Abelian classical r-matrices},''
  \href{http://dx.doi.org/10.1088/1751-8113/49/44/445403}{{\em J. Phys.}
  {\bfseries A49} no.~44, (2016) 445403},
\href{http://arxiv.org/abs/1607.00795}{{\ttfamily arXiv:1607.00795 [hep-th]}}.

\bibitem{vanTongeren:2016eeb}
S.~J. van Tongeren, ``{Almost abelian twists and AdS/CFT},''
  \href{http://dx.doi.org/10.1016/j.physletb.2016.12.002}{{\em Phys. Lett. B}
  {\bfseries 765} (2017) 344--351},
  \href{http://arxiv.org/abs/1610.05677}{{\ttfamily arXiv:1610.05677
  [hep-th]}}.

\bibitem{Araujo:2017jkb}
T.~Araujo, I.~Bakhmatov, E.~\'O~Colg\'ain, J.~Sakamoto, M.~M. Sheikh-Jabbari,
  and K.~Yoshida, ``{Yang-Baxter $\sigma$-models, conformal twists, and
  noncommutative Yang-Mills theory},''
  \href{http://dx.doi.org/10.1103/PhysRevD.95.105006}{{\em Phys. Rev.}
  {\bfseries D95} no.~10, (2017) 105006},
\href{http://arxiv.org/abs/1702.02861}{{\ttfamily arXiv:1702.02861 [hep-th]}}.

\bibitem{Hong:2018tlp}
M.~Hong, Y.~Kim, and E.~\'O~Colg\'ain, ``{On non-Abelian T-duality for
  non-semisimple groups},''
  \href{http://dx.doi.org/10.1140/epjc/s10052-018-6502-9}{{\em Eur. Phys. J.}
  {\bfseries C78} no.~12, (2018) 1025},
\href{http://arxiv.org/abs/1801.09567}{{\ttfamily arXiv:1801.09567 [hep-th]}}.

\bibitem{Borsato:2018idb}
R.~Borsato and L.~Wulff, ``{Non-abelian T-duality and Yang-Baxter deformations
  of Green-Schwarz strings},''
  \href{http://dx.doi.org/10.1007/JHEP08(2018)027}{{\em JHEP} {\bfseries 08}
  (2018) 027},
\href{http://arxiv.org/abs/1806.04083}{{\ttfamily arXiv:1806.04083 [hep-th]}}.

\bibitem{Bakhmatov:2018bvp}
I.~Bakhmatov and E.~T. Musaev, ``{Classical Yang-Baxter equation from
  $\beta$-supergravity},''
  \href{http://dx.doi.org/10.1007/JHEP01(2019)140}{{\em JHEP} {\bfseries 01}
  (2019) 140},
\href{http://arxiv.org/abs/1811.09056}{{\ttfamily arXiv:1811.09056 [hep-th]}}.

\bibitem{Bakhmatov:2020kul}
I.~Bakhmatov, K.~Gubarev, and E.~T. Musaev, ``{Non-abelian tri-vector
  deformations in $d=11$ supergravity},''
  \href{http://dx.doi.org/10.1007/JHEP05(2020)113}{{\em JHEP} {\bfseries 05}
  (2020) 113}, \href{http://arxiv.org/abs/2002.01915}{{\ttfamily
  arXiv:2002.01915 [hep-th]}}.

\bibitem{Bakhmatov:2019dow}
I.~Bakhmatov, N.~S. Deger, E.~T. Musaev, E.~\'O~Colg\'ain, and M.~M.
  Sheikh-Jabbari, ``{Tri-vector deformations in $d=11$ supergravity},''
  \href{http://dx.doi.org/10.1007/JHEP08(2019)126}{{\em JHEP} {\bfseries 08}
  (2019) 126},
\href{http://arxiv.org/abs/1906.09052}{{\ttfamily arXiv:1906.09052 [hep-th]}}.

\bibitem{Peeters:2007wn}
K.~Peeters, ``Introducing Cadabra: A Symbolic computer algebra system for field
  theory problems,''
\href{http://arxiv.org/abs/hep-th/0701238}{{\ttfamily arXiv:hep-th/0701238
  [hep-th]}}.

\bibitem{gengensugra_2022}
I.~Bakhmatov, A.~\c{C}atal \"Ozer, N.~S. Deger, K.~Gubarev, and E.~T. Musaev,
  ``{GenGenSugra GitHub},'' 8, 2022.
\newblock \url{https://github.com/emusaev/gengensugra/wiki}.

\bibitem{Duff:1989tf}
M.~Duff, ``{Duality rotations in string theory},''
  \href{http://dx.doi.org/10.1016/0550-3213(90)90520-N}{{\em Nucl.Phys.}
  {\bfseries B335} (1990) 610}.

\bibitem{Tseytlin:1990va}
A.~A. Tseytlin, ``{Duality symmetric closed string theory and interacting
  chiral scalars},'' \href{http://dx.doi.org/10.1016/0550-3213(91)90266-Z}{{\em
  Nucl.Phys.} {\bfseries B350} (1991) 395--440}.

\bibitem{Tseytlin:1990nb}
A.~A. Tseytlin, ``{Duality symmetric formulation of string world sheet
  dynamics},'' \href{http://dx.doi.org/10.1016/0370-2693(90)91454-J}{{\em
  Phys.Lett.} {\bfseries B242} (1990) 163--174}.

\bibitem{Siegel:1993th}
W.~Siegel, ``{Superspace duality in low-energy superstrings},''
  \href{http://dx.doi.org/10.1103/PhysRevD.48.2826}{{\em Phys.Rev.} {\bfseries
  D48} (1993) 2826--2837},
\href{http://arxiv.org/abs/hep-th/9305073}{{\ttfamily arXiv:hep-th/9305073
  [hep-th]}}.

\bibitem{Siegel:1993xq}
W.~Siegel, ``{Two vierbein formalism for string inspired axionic gravity},''
  \href{http://dx.doi.org/10.1103/PhysRevD.47.5453}{{\em Phys. Rev.} {\bfseries
  D47} (1993) 5453--5459},
\href{http://arxiv.org/abs/hep-th/9302036}{{\ttfamily arXiv:hep-th/9302036
  [hep-th]}}.

\bibitem{Fradkin:1984ai}
E.~Fradkin and A.~A. Tseytlin, ``{Quantum Equivalance of Dual Field
  Theories},''
\href{http://dx.doi.org/10.1016/0003-4916(85)90225-8}{{\em Annals Phys.}
  {\bfseries 162} (1985) 31}.

\bibitem{Tseytlin:1990ar}
A.~A. Tseytlin and P.~C. West, ``{Two Remarks on Chiral Scalars},''
\href{http://dx.doi.org/10.1103/PhysRevLett.65.541}{{\em Phys.Rev.Lett.}
  {\bfseries 65} (1990) 541--542}.

\bibitem{Hull:2009mi}
C.~Hull and B.~Zwiebach, ``{Double field theory},''
  \href{http://dx.doi.org/10.1088/1126-6708/2009/09/099}{{\em JHEP} {\bfseries
  0909} (2009) 099}, \href{http://arxiv.org/abs/0904.4664}{{\ttfamily
  arXiv:0904.4664 [hep-th]}}.

\bibitem{Hull:2009zb}
C.~Hull and B.~Zwiebach, ``{The Gauge algebra of double field theory and
  Courant brackets,},''
  \href{http://dx.doi.org/10.1088/1126-6708/2009/09/090}{{\em JHEP} {\bfseries
  {\bf 0909}} (2009) 090},
\href{http://arxiv.org/abs/0908.1792}{{\ttfamily arXiv:0908.1792 [hep-th]}}.

\bibitem{Hohm:2010jy}
O.~Hohm, C.~Hull, and B.~Zwiebach, ``{Background independent action for double
  field theory},'' \href{http://dx.doi.org/10.1007/JHEP07(2010)016}{{\em JHEP}
  {\bfseries 1007} (2010) 016},
\href{http://arxiv.org/abs/1003.5027}{{\ttfamily arXiv:1003.5027 [hep-th]}}.

\bibitem{Hohm:2010pp}
O.~Hohm, C.~Hull, and B.~Zwiebach, ``{Generalized metric formulation of double
  field theory},'' \href{http://dx.doi.org/10.1007/JHEP08(2010)008}{{\em JHEP}
  {\bfseries 1008} (2010) 008},
  \href{http://arxiv.org/abs/1006.4823}{{\ttfamily arXiv:1006.4823 [hep-th]}}.

\bibitem{Hohm:2011dv}
O.~Hohm, S.~K. Kwak, and B.~Zwiebach, ``{Double Field Theory of Type II
  Strings},'' \href{http://dx.doi.org/10.1007/JHEP09(2011)013}{{\em JHEP}
  {\bfseries 09} (2011) 013}, \href{http://arxiv.org/abs/1107.0008}{{\ttfamily
  arXiv:1107.0008 [hep-th]}}.

\bibitem{Geissbuhler:2013uka}
D.~Geissbuhler, D.~Marques, C.~Nunez, and V.~Penas, ``{Exploring Double Field
  Theory},'' \href{http://dx.doi.org/10.1007/JHEP06(2013)101}{{\em JHEP}
  {\bfseries 06} (2013) 101}, \href{http://arxiv.org/abs/1304.1472}{{\ttfamily
  arXiv:1304.1472 [hep-th]}}.

\bibitem{Hull:2014mxa}
C.~M. Hull, ``{Finite Gauge Transformations and Geometry in Double Field
  Theory},'' \href{http://dx.doi.org/10.1007/JHEP04(2015)109}{{\em JHEP}
  {\bfseries 04} (2015) 109}, \href{http://arxiv.org/abs/1406.7794}{{\ttfamily
  arXiv:1406.7794 [hep-th]}}.

\bibitem{Sazdovic:2017lqo}
B.~Sazdovi\'c, ``{Open string T-duality in double space},''
  \href{http://dx.doi.org/10.1140/epjc/s10052-017-5190-1}{{\em Eur. Phys. J. C}
  {\bfseries 77} no.~9, (2017) 634},
  \href{http://arxiv.org/abs/1704.01163}{{\ttfamily arXiv:1704.01163
  [hep-th]}}.

\bibitem{Berman:2012vc}
D.~S. Berman, M.~Cederwall, A.~Kleinschmidt, and D.~C. Thompson, ``{The gauge
  structure of generalised diffeomorphisms},''
  \href{http://dx.doi.org/10.1007/JHEP01(2013)064}{{\em JHEP} {\bfseries 1301}
  (2013) 064},
\href{http://arxiv.org/abs/1208.5884}{{\ttfamily arXiv:1208.5884 [hep-th]}}.

\bibitem{Sakatani:2016fvh}
Y.~Sakatani, S.~Uehara, and K.~Yoshida, ``{Generalized gravity from modified
  DFT},'' \href{http://dx.doi.org/10.1007/JHEP04(2017)123}{{\em JHEP}
  {\bfseries 04} (2017) 123},
\href{http://arxiv.org/abs/1611.05856}{{\ttfamily arXiv:1611.05856 [hep-th]}}.

\bibitem{Catal-Ozer:2019tmm}
A.~\c{C}atal{-}{\"O}zer and S.~Tunal{\i}, ``{Yang-Baxter Deformation as an
  O(d,d) Transformation},''
  \href{http://dx.doi.org/10.1088/1361-6382/ab6f7e}{{\em Class. Quant. Grav.}
  {\bfseries 37} no.~7, (2020) 075003},
  \href{http://arxiv.org/abs/1906.09053}{{\ttfamily arXiv:1906.09053
  [hep-th]}}.

\bibitem{Baguet:2016prz}
A.~Baguet, M.~Magro, and H.~Samtleben, ``{Generalized IIB supergravity from
  exceptional field theory},''
  \href{http://dx.doi.org/10.1007/JHEP03(2017)100}{{\em JHEP} {\bfseries 03}
  (2017) 100},
\href{http://arxiv.org/abs/1612.07210}{{\ttfamily arXiv:1612.07210 [hep-th]}}.

\bibitem{Ciceri:2016dmd}
F.~Ciceri, A.~Guarino, and G.~Inverso, ``{The exceptional story of massive IIA
  supergravity},'' \href{http://dx.doi.org/10.1007/JHEP08(2016)154}{{\em JHEP}
  {\bfseries 08} (2016) 154}, \href{http://arxiv.org/abs/1604.08602}{{\ttfamily
  arXiv:1604.08602 [hep-th]}}.

\bibitem{Berman:2010is}
D.~S. Berman and M.~J. Perry, ``{Generalized Geometry and M theory},''
  \href{http://dx.doi.org/10.1007/JHEP06(2011)074}{{\em JHEP} {\bfseries 06}
  (2011) 074}, \href{http://arxiv.org/abs/1008.1763}{{\ttfamily arXiv:1008.1763
  [hep-th]}}.

\bibitem{Berman:2011jh}
D.~S. Berman, H.~Godazgar, M.~J. Perry, and P.~West, ``{Duality Invariant
  Actions and Generalised Geometry},''
  \href{http://dx.doi.org/10.1007/JHEP02(2012)108}{{\em JHEP} {\bfseries 1202}
  (2012) 108},
\href{http://arxiv.org/abs/1111.0459}{{\ttfamily arXiv:1111.0459 [hep-th]}}.

\bibitem{Berman:2011pe}
D.~S. Berman, H.~Godazgar, and M.~J. Perry, ``{SO(5,5) duality in M-theory and
  generalized geometry},''
  \href{http://dx.doi.org/10.1016/j.physletb.2011.04.046}{{\em Phys.Lett.}
  {\bfseries B700} (2011) 65--67},
\href{http://arxiv.org/abs/1103.5733}{{\ttfamily arXiv:1103.5733 [hep-th]}}.

\bibitem{Hohm:2013jma}
O.~Hohm and H.~Samtleben, ``{U-duality covariant gravity},''
  \href{http://dx.doi.org/10.1007/JHEP09(2013)080}{{\em JHEP} {\bfseries 1309}
  (2013) 080},
\href{http://arxiv.org/abs/1307.0509}{{\ttfamily arXiv:1307.0509 [hep-th]}}.

\bibitem{Hohm:2013pua}
O.~Hohm and H.~Samtleben, ``{Exceptional Form of D=11 Supergravity},''
  \href{http://dx.doi.org/10.1103/PhysRevLett.111.231601}{{\em Phys.Rev.Lett.}
  {\bfseries 111} (2013) 231601},
\href{http://arxiv.org/abs/1308.1673}{{\ttfamily arXiv:1308.1673 [hep-th]}}.

\bibitem{Hohm:2013vpa}
O.~Hohm and H.~Samtleben, ``{Exceptional Field Theory I: $E_{6(6)}$ covariant
  Form of M-Theory and Type IIB},''
  \href{http://dx.doi.org/10.1103/PhysRevD.89.066016}{{\em Phys.Rev.}
  {\bfseries D89} (2014) 066016},
\href{http://arxiv.org/abs/1312.0614}{{\ttfamily arXiv:1312.0614 [hep-th]}}.

\bibitem{Musaev:2014lna}
E.~Musaev and H.~Samtleben, ``{Fermions and supersymmetry in E$_{6(6)}$
  exceptional field theory},''
  \href{http://dx.doi.org/10.1007/JHEP03(2015)027}{{\em JHEP} {\bfseries 1503}
  (2015) 027},
\href{http://arxiv.org/abs/1412.7286}{{\ttfamily arXiv:1412.7286 [hep-th]}}.

\bibitem{Abzalov:2015ega}
A.~Abzalov, I.~Bakhmatov, and E.~T. Musaev, ``{Exceptional field theory:
  $SO(5,5)$},'' \href{http://dx.doi.org/10.1007/JHEP06(2015)088}{{\em JHEP}
  {\bfseries 06} (2015) 088},
\href{http://arxiv.org/abs/1504.01523}{{\ttfamily arXiv:1504.01523 [hep-th]}}.

\bibitem{Musaev:2015ces}
E.~T. Musaev, ``{Exceptional field theory: $SL(5)$},''
  \href{http://dx.doi.org/10.1007/JHEP02(2016)012}{{\em JHEP} {\bfseries 02}
  (2016) 012},
\href{http://arxiv.org/abs/1512.02163}{{\ttfamily arXiv:1512.02163 [hep-th]}}.

\bibitem{Coimbra:2011ky}
A.~Coimbra, C.~Strickland-Constable, and D.~Waldram, ``{$E_{d(d)} \times
  \mathbb{R}^+$ generalised geometry, connections and M theory},''
  \href{http://dx.doi.org/10.1007/JHEP02(2014)054}{{\em JHEP} {\bfseries 1402}
  (2014) 054},
\href{http://arxiv.org/abs/1112.3989}{{\ttfamily arXiv:1112.3989 [hep-th]}}.

\bibitem{Berman:2012uy}
D.~S. Berman, E.~T. Musaev, and D.~C. Thompson, ``{Duality Invariant M-theory:
  Gauged supergravities and Scherk-Schwarz reductions},''
  \href{http://dx.doi.org/10.1007/JHEP10(2012)174}{{\em JHEP} {\bfseries 1210}
  (2012) 174},
\href{http://arxiv.org/abs/1208.0020}{{\ttfamily arXiv:1208.0020 [hep-th]}}.

\bibitem{Blair:2014zba}
C.~D.~A. Blair and E.~Malek, ``{Geometry and fluxes of SL(5) exceptional field
  theory},'' \href{http://dx.doi.org/10.1007/JHEP03(2015)144}{{\em JHEP}
  {\bfseries 03} (2015) 144},
\href{http://arxiv.org/abs/1412.0635}{{\ttfamily arXiv:1412.0635 [hep-th]}}.

\bibitem{Grana:2012rr}
M.~Gra{\~n}a and D.~Marqu{\'e}s, ``{Gauged Double Field Theory},''
  \href{http://dx.doi.org/10.1007/JHEP04(2012)020}{{\em JHEP} {\bfseries 1204}
  (2012) 020},
\href{http://arxiv.org/abs/1201.2924}{{\ttfamily arXiv:1201.2924 [hep-th]}}.

\bibitem{Wulff:2018aku}
L.~Wulff, ``{Trivial solutions of generalized supergravity vs non-abelian
  T-duality anomaly},''
  \href{http://dx.doi.org/10.1016/j.physletb.2018.04.025}{{\em Phys. Lett. B}
  {\bfseries 781} (2018) 417--422},
  \href{http://arxiv.org/abs/1803.07391}{{\ttfamily arXiv:1803.07391
  [hep-th]}}.

\bibitem{Thompson:2011uw}
D.~C. Thompson, ``{Duality Invariance: From M-theory to Double Field Theory},''
  \href{http://dx.doi.org/10.1007/JHEP08(2011)125}{{\em JHEP} {\bfseries 1108}
  (2011) 125},
\href{http://arxiv.org/abs/1106.4036}{{\ttfamily arXiv:1106.4036 [hep-th]}}.

\end{thebibliography}\endgroup
\bibliographystyle{utphys.bst}

\end{document}